\documentclass[rog]{agutex}

\usepackage{graphicx}
\usepackage{amssymb}
\usepackage{wasysym}

\authorrunninghead{FEULNER}

\titlerunninghead{THE FAINT YOUNG SUN PROBLEM}

\authoraddr{Georg Feulner, Potsdam Institute for Climate Impact
  Research, P.O.\ Box 60 12 03, D--14412 Potsdam, Germany
  (feulner@pik-potsdam.de)}

\newcommand{\amm}{NH$_3$}
\newcommand{\ntwo}{N$_2$}
\newcommand{\cotwo}{CO$_2$}
\newcommand{\methane}{CH$_4$}
\newcommand{\water}{H$_2$O}
\newcommand{\htwos}{H$_2$S}
\newcommand{\otwo}{O$_2$}
\newcommand{\siderite}{FeCO$_3$}
\newcommand{\magnetite}{Fe$_3$O$_4$}

\newcommand{\tgyr}{Tg~yr$^{-1}$}

\begin{document}

%
%

\title{The faint young Sun problem}
%

%
%

\authors{Georg Feulner \altaffilmark{1}}

\altaffiltext{1}{Earth System Analysis, Potsdam Institute for Climate
  Impact Research, Potsdam, Germany}








%
%


\begin{abstract}

  For more than four decades, scientists have been trying to find an
  answer to one of the most fundamental questions in paleoclimatology,
  the `faint young Sun problem'. For the early Earth, models of
  stellar evolution predict a solar energy input to the climate system
  which is about 25\% lower than today. This would result in a
  completely frozen world over the first two billion years in the
  history of our planet, if all other parameters controlling Earth's
  climate had been the same. Yet there is ample evidence for the
  presence of liquid surface water and even life in the Archean (3.8
  to 2.5 billion years before present), so some effect (or effects)
  must have been compensating for the faint young Sun. A wide range of
  possible solutions have been suggested and explored during the last
  four decades, with most studies focussing on higher concentrations
  of atmospheric greenhouse gases like carbon dioxide, methane or
  ammonia. All of these solutions present considerable difficulties,
  however, so the faint young Sun problem cannot be regarded as
  solved. Here I review research on the subject, including the latest
  suggestions for solutions of the faint young Sun problem and recent
  geochemical constraints on the composition of Earth's early
  atmosphere. Furthermore, I will outline the most promising
  directions for future research. In particular I would argue that
  both improved geochemical constraints on the state of the Archean
  climate system and numerical experiments with state-of-the-art
  climate models are required to finally assess what kept the oceans
  on the Archean Earth from freezing over completely.

\end{abstract}

%
%

%

\begin{article}

%
%

\section{Introduction}
\label{s:intro}

The faint young Sun problem for Earth's early climate has been briefly
reviewed a few times in the past, for example in the general context
of climate change on geological timescales \citep{Crowley1983,
  Barron1984}, the formation and early history of Earth
\citep{Zahnle2007}, the evolution of Earth's atmosphere and climate
\citep{Pollack1991, Kasting1993, Shaw2008, Nisbet2011}, life on the
early Earth \citep{Nisbet2001}, evolution of the terrestrial planets
and considerations of planetary habitability \citep{Pollack1979,
  Rampino1994, Kasting2003} or the evolution of the Sun
\citep{KastingGrinspoon1991, Guedel2007}. The more comprehensive
reviews of this topic are somewhat dated by now, however, and most
look at the issue from the point of view of the global energy balance
without exploring important internal aspects of the climate system
like the transport of heat.

This paper presents a new and detailed review of the faint young Sun
problem and is organized as follows. Section~\ref{s:problem} describes
the evidence for a faint young Sun and for the existence of liquid
water on early Earth. Section~\ref{s:solutions} explores in what ways
the faint young Sun problem could be solved in principle before the
options are discussed in detail in the following
sections. Section~\ref{s:massloss} looks at modifications of the
standard solar model, in particular the possibility of a strong
mass-loss of the young Sun. The most likely solution of the faint
young Sun problem in terms of an enhanced greenhouse effect is
discussed in Section~\ref{s:greenhouse}, the main Section of this
review paper. Then the effects of clouds (Section~\ref{s:clouds}) and
differences in rotation rate and continental configuration
(Section~\ref{s:rotcont}) will be explored, before the review is
concluded by a summary and suggestions for future research in
Section~\ref{s:disc}.

\section{The Faint Young Sun Problem}
\label{s:problem}

In this Section, the faint young Sun problem is introduced, beginning
with a discussion of the evolution of the Sun on long timescales.

\subsection{A Fainter Sun in the Past}
\label{s:sun}

By the 1950s, stellar astrophysicists had worked out the physical
principles governing the structure and evolution of stars
\citep{Kippenhahn1994}. This allowed the construction of theoretical
models for the stellar interior and the evolutionary changes occurring
during the lifetime of a star. Applying these principles to the Sun,
it became clear that the luminosity of the Sun had to change over
time, with the young Sun being considerably less luminous than today
\citep{Hoyle1958, Schwarzschild1958}.

\begin{figure*}
\centerline{\includegraphics[width=10cm]{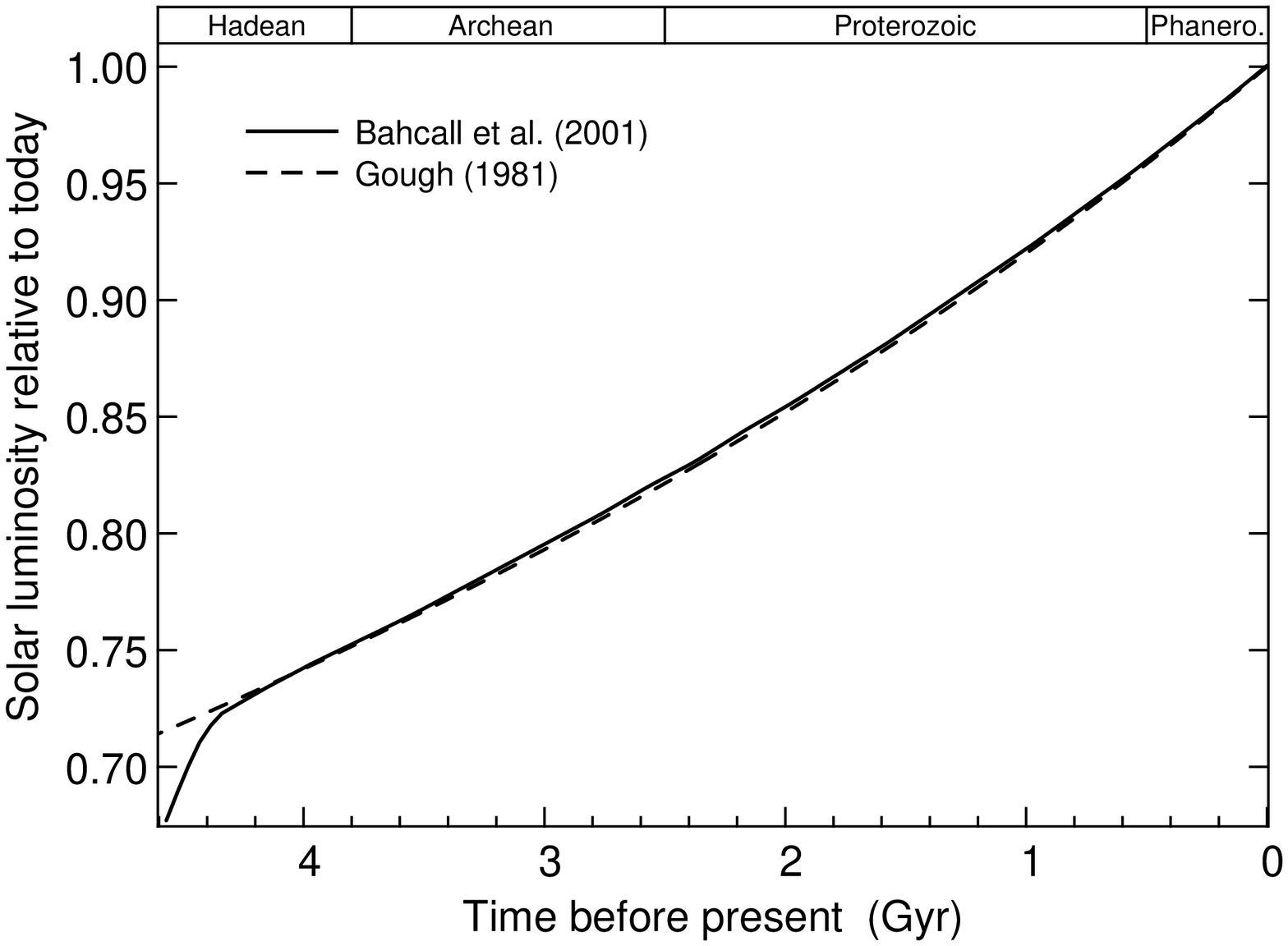}}
\caption{Evolution of solar luminosity over the four geologic eons for
  the standard solar model described in \citet[][\textit{solid
    line}]{Bahcall2001} and according to the approximation formula
  \citep[][\textit{dashed line}]{Gough1981} given in
  equation~(\ref{e:sollum}).}
\label{f:sollum}
\end{figure*}

According to standard solar models, when nuclear fusion ignited in the
core of the Sun at the time of its arrival on what is called the
zero-age main sequence (ZAMS) 4.57~Ga (1~Ga = $10^9$ years ago), the
bolometric luminosity of the Sun (the solar luminosity integrated over
all wavelengths) was about 30\% lower as compared to the present epoch
\citep{Newman1977}. The long-term evolution of the bolometric solar
luminosity $L (t)$ as a function of time $t$ can be approximated by a
simple formula \citep{Gough1981}

\begin{equation}
  \label{e:sollum}
  \frac{L \, (t)}{L_\odot} \: = \: \frac{1}{1 + \frac{2}{5} \left( 1 -
    \frac{t}{t_\odot} \right) } ,
\end{equation}

where $L_\odot = 3.85 \times 10^{26}$~W is the present-day solar
luminosity and $t_\odot = 4.57$~Gyr (1~Gyr = $10^9$~years) is the age
of the Sun. Except for the first $\sim 0.2$~Gyr in the life of the
young Sun, this approximation agrees very well with the time evolution
calculated with more recent standard solar models
\citep[e.g.,][]{Bahcall2001}, see the comparison in
Figure~\ref{f:sollum}.

Note that solar models had been under intense scrutiny for a long time
in the context of the ``solar neutrino problem'', an apparent
deficiency of neutrinos observed in terrestrial neutrino detectors
\citep{Haxton1995} which is now considered to be resolved by a
modification of the standard model of particle physics
\citep{Mohapatra2006} rather than to be an indication of problems with
solar models. Furthermore, the time evolution of the Sun's luminosity
has been shown to be a very robust feature of solar models
\citep{Newman1977, Bahcall2001}. Thus it appears highly unlikely that
the prediction of low luminosity for the early Sun is due to
fundamental problems with solar models. (Slightly modified solar
models involving a larger mass loss in the past will be discussed in
Section~\ref{s:massloss}.)

In a way the robustness of the luminosity evolution of stellar models
is not surprising, since the gradual rise in solar luminosity is a
simple physical consequence of the way the Sun generates energy by
nuclear fusion of hydrogen to helium in its core. Over time, Helium
nuclei accumulate, increasing the mean molecular weight within the
core. For a stable, spherical distribution of mass twice the total
kinetic energy is equal to the absolute value of the potential
energy. According to this virial theorem, the Sun's core contracts and
heats up to keep the star stable, resulting in a higher energy
conversion rate and hence a higher luminosity. There seems no
possibility for escape \citep{Gough1981}: ``The gradual increase in
luminosity during the core hydrogen burning phase of evolution of a
star is an inevitable consequence of Newtonian physics and the
functional dependence of the thermonuclear reaction rates on density,
temperature and composition.''

In addition to this slow evolution of the bolometric solar luminosity
over timescales of $\sim 10^9$~yr, the Sun exhibits variability on
shorter timescales of up to $\sim 10^3$~yr \citep{Frohlich2004}. This
variability in solar radiation is a manifestation of changes in its
magnetic activity related to the solar magnetic field created by a
magnetohydrodynamic dynamo within the Sun \citep{Weiss2000}. The
bolometric solar luminosity is dominated by radiation in the visible
spectral range originating from the Sun's lower atmosphere which shows
very little variation with solar activity \citep{Frohlich2004}. For
the present-day Sun, for example, total solar irradiance varies by
only $\simeq 0.1$\% over the 11-year sunspot cycle \citep{Gray2010}.

The Sun's ultraviolet radiation, on the other hand, is predominantly
emitted by the hotter upper layers of the solar atmosphere which are
subject to much larger variability \citep{Lean1987,
  Frohlich2004}. Solar variability (and thus ultraviolet luminosity)
was higher in the past due to a steady decrease in magnetic activity
over time caused by the gradual slowing of the Sun's rotation which
ultimately drives the magnetohydrodynamic dynamo \citep{Zahnle1982,
  Dorren1994, Guedel2007}. From observations of young stars similar to
the Sun one can infer a decrease in rotation rate $\Omega_\odot$ of
the Sun with time $t$ which follows a power law

\begin{equation}
\Omega_\odot \: \propto \: t^{-0.6}
\end{equation}

\citep{Guedel2007}. For the same reason, the solar wind was stronger
for the young Sun, with consequences for the early Earth's
magnetosphere and the loss of volatiles and water from the early
atmosphere \citep{Sterenborg2011}, especially considering the fact
that the strength of Earth's magnetic field was estimated to be $\sim
50 - 70$\% of the present-day field strength $3.4-3.45$~Ga
\citep{Tarduno2010}. The effects of these changes in ultraviolet
radiation and solar wind will be briefly discussed later on.

Coming back to the lower bolometric luminosity of the Sun, an estimate
of the amount of radiative forcing of the climate system this
reduction corresponds to is given by $\Delta F = \Delta S_0 (1-A)/4$
(the change in incoming solar radiation corrected for geometry and
Earth's albedo $A$). Using the present-day solar constant $S_0 \simeq
1361$~W~m$^{-2}$ \citep{Kopp2011} and Earth's current albedo $A \simeq
0.3$ yields values of $\Delta F \approx 60$~W~m$^{-2}$ and $\Delta F
\approx 40$~W~m$^{-2}$ at times 3.8~Ga and 2.5~Ga, respectively. For
comparison, the net anthropogenic radiative forcing in 2005 is
estimated to be $\simeq 1.6$~W~m$^{-2}$ \citep{IPCC-4-1-2}.

Solar physicists speculated early on that this large reduction of the
incoming solar radiation might have had consequences for the evolution
of Earth's climate \citep{Schwarzschild1958}: ``Can this change in the
brightness of the sun have had some geophysical or geological
consequences that might be detectable?''

\subsection{Evidence for Liquid Water on Early Earth}
\label{s:water}

A few years later, the possible consequences of these astrophysical
findings of a faint young Sun on the climate of Earth were first
discussed by \citet{Ringwood1961}, who pointed out that ``[o]ther
factors being equal, [\dots] the surface of the earth during the
period between its birth, 4.5 billion years ago, and 3 billion years
ago, would have passed through an intense ice age.''

A significant reduction in solar energy input can result in dramatic
effects for the Earth's climate due to the ice-albedo feedback:
Decreasing temperatures result in larger areas covered in ice which
has a large albedo and thus reflects more radiation back into space,
further enhancing the cooling.  Climate models show the importance of
this ice-albedo feedback for the Earth's global energy balance: Once a
critical luminosity threshold is reached, this results in run-away
glaciation and completely ice-covered oceans, a ``snowball Earth''
\citep{Kirschvink1992} state (see also Figure~\ref{f:iceline} and the
discussion in Section~\ref{s:rotcont}). It should be noted that a
recent modeling study suggests a third stable state in which a narrow
strip in the tropics remains free of ice due to the combined effects
of the lower albedo of snow-free sea ice and the reduced cloud cover
in this region \citep{Abbot2011}.

While earlier models placed the critical luminosity threshold at
$2-5$\% below the present-day value for modern continental
configuration \citep{Budyko1969, Sellers1969, Gerard1992}, later
studies with more sophisticated models found values of $10-15$\% and
up to $18$\% for global ocean conditions \citep{Jenkins1993b,
  Longdoz1997}. Differences in critical luminosity between
energy-balance models can be attributed to the sensitivity of the ice
line to the parametrization of meridional heat transport
\citep{Held1975, Lindzen1977, Ikeda1999}, to geography
\citep{Crowley1993} and to the question whether the climate model is
coupled to a dynamic ice-sheet model or not
\citep{Hyde2000}. Furthermore, the position of the ice line in
simulations with comprehensive general circulation models is strongly
influenced by ocean dynamics \citep{Poulsen2001}.

Once snowball Earth conditions are reached, it requires high
concentrations of greenhouse gases in the atmosphere (for example from
the gradual build-up of volcanic carbon dioxide in the atmosphere) to
return to a warmer climate state due to the high reflectivity of the
ice, although volcanic ash and material from meteorite impacts might
lower the albedo and thus increase the absorption of solar radiation
\citep{Schatten1982}.

Note that the oceans would not have been frozen completely (i.e., down
to the ocean floor) because of the flow of geothermal heat from the
Earth's interior. For the case of a cold climate on early Earth, the
thickness of the ice layer at the oceans' surface has been estimated
with a simple one-dimensional heat flow model to be a few hundred
meters given the higher geothermal heat flux at that time
\citep{Bada1994}. Models like this ignore the effects of ice dynamics,
however.

Contrary to these expected climatic effects of the faint young Sun,
however, there is ample evidence for the presence of liquid water at
the surface of the young Earth during the Hadean and Archean eons. For
the purpose of this review, the Hadean eon is defined to span the
period from the Earth's formation 4.56 Ga to 3.8~Ga and the Archean
eon is assumed to last from the end of the Hadean to 2.5~Ga. Note that
the Hadean is not officially defined and that there is no agreement
about the Hadean-Archean boundary which is frequently set at 4.0~Ga
\citep[see, e.g.,][for discussions]{Zahnle2007, Goldblatt2010}. It
appears logical, however, to define the beginning of the Archean at
the end of the period of intense impacts from space known as the `Late
Heavy Bombardment' \citep{Tera1974, Wetherill1975, Hartmann2000,
  Kring2002} occurring $\sim 4.0-3.8$~Ga, although the exact end of
that period is not resolved in the geological record. Irrespective of
these matters of definition, it is important to realize that the
Archean, the main focus of this review, spans a very long period of
time in the history of Earth.

Tentative evidence for liquid water on the early Earth can be found in
the Hadean. No rocks are known from the Hadean due to the exponential
decrease of preservation with age, yet some information on the surface
conditions during those earlier times can be derived from the mineral
zircon (ZrSiO$_4$) preserved from the Hadean in younger rocks
\citep{Harrison2009}. Indeed, zircon grains may provide evidence for
liquid water even before the Archean, as early as 4.2~Ga
\citep{Mojzsis2001, Wilde2001, Valley2002, Harrison2009}.

Note, however, that the environment in which this Hadean ocean existed
was considerably different from the Archean \citep[for a review of the
following outline of events see, e.g.,][]{Zahnle2007}.  The Earth was
formed by gravitational accretion of smaller bodies (planetesimals)
formed in the nebula surrounding the young Sun. The large impact
forming the Moon occurred after $~ 50$~Myr towards the end of the
accretion period. After this event, Earth was enshrouded in rock vapor
for $~1000$~yr. A strong greenhouse effect (caused by large amounts of
carbon dioxide and water vapor degassing from the mantle) and tidal
heating by the still tightly-orbiting Moon kept the surface covered by
a magma ocean for a few million years after the Moon-forming
impact. Then the crust solidified and a hot water ocean with
temperatures of $\sim 500$~K formed under a dense atmosphere
containing $\sim 100$~bar of carbon dioxide. The carbon dioxide in the
atmosphere was then subducted into the mantle over timescales of
$10^{7-8}$~yr, before the Late Heavy Bombardment ($\sim 4.0-3.8$~Ga)
set the stage for the Archean eon. It is thus clear that the processes
resulting in a liquid-water ocean in the Hadean are considerably
different from the Archean, so they will not be discussed further in
this review.

Geologic evidence for liquid surface water during the Archean is
mostly based on sedimentary rock laid down in a variety of aqueous
conditions up to 3.5~Ga and possibly as early as 3.8~Ga, and there is
no evidence for wide-spread glaciations during the entire Archean [see
\citealt{Lowe1980, Walker1982, Walker1983}; for more recent overviews
of Archean geology in general see, e.g., \citealt{Fowler2002,
  Eriksson2004, Benn2006}]. Telltale signs of liquid water include
pillow lavas which are formed when lava extrudes under water, ripple
marks resulting from sediment deposition under the influence of waves,
and mud cracks.

Furthermore, there is evidence for microbial life in the Archean
derived from microfossils or stromatolites/microbial mats in rocks of
ages between 2.5 and 3.5~Gyr \citep{Barghoorn1966, Altermann2003,
  Schopf2006}. Although all life on Earth is based on the existence of
liquid water \citep[e.g.,][]{Pace2001}, the mere existence of life is
only a poor contraint on ice cover. The early evidence of
photosynthetic cyanobacteria and stromatolites, however, constitutes
further evidence for an early Earth not permanently covered by ice (or
at least for continuously ice-free regions in the oceans). One could,
in principle, imagine photosynthetic life under a thin ice cover in
the tropics of a snowball Earth as postulated by \citet{McKay2000} and
investigated in \citet{Pollard2005, Pollard2006}. Later studies have
indicated, however, that ice cover would have been too thick even in
the tropics \citep{Warren2002, Goodman2006, Warren2006}, making such a
scenario unlikely.

In summary, there are multiple lines of independent evidence
suggesting the existence of liquid water on Earth's surface during the
Archean, when the Sun was considerably fainter than today.

\subsection{Temperatures during the Archean}
\label{s:temphist}

\begin{figure*}
\centerline{\includegraphics[width=10cm]{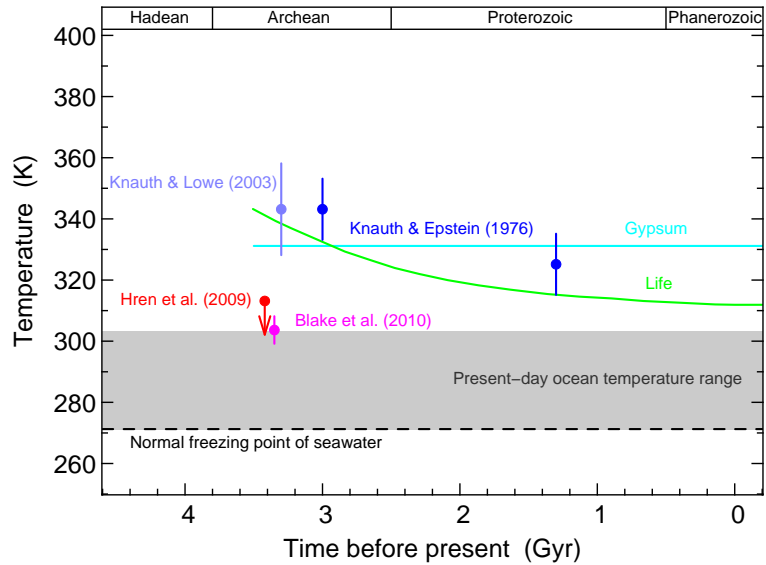}}
\caption{Constraints on ocean temperatures during the Archean. The
  existence of diverse life since about 3.5~Gyr and the typical ranges
  of temperature tolerance of living organisms suggests the upper
  limit indicated by the \textit{green line}
  \citep{Walker1982}. Evaporate minerals are present since about
  3.5~Gyr, and the fact that many were initially deposited as gypsum
  sets an upper limit at 58$^\circ$C \textit{(cyan line)}
  \citep{Holland1978}. The comparatively high (but controversial, see
  the text for discussion) temperatures derived from oxygen isotope
  ratios in cherts are shown in \textit{blue} \citep{Knauth1976,
    Knauth2003}. More recent estimates based on a combination of
  oxygen and hydrogen isotope ratios \citep{Hren2009} and the oxygen
  isotope composition of phosphates \citep{Blake2010} are shown in
  \textit{red} and \textit{magenta}, respectively. The range of
  present-day ocean temperatures is indicated in \textit{gray}
  \citep{WOA2009_Temp}, the freezing point of seawater at normal
  pressure and for present-day salinity by the \textit{dashed
    line}. Modified and updated after \citet{Walker1982}.}
\label{f:temphist}
\end{figure*}

It is one of the key characteristics of water that it remains liquid
over a rather wide range of temperatures, so the question arises of
how warm the Archean climate actually was. The constraints on and
estimates of Archean ocean temperatures discussed below are summarized
in Figure~\ref{f:temphist}.

Upper limits to Archean climate temperatures can be mainly derived
from two lines of argument. First, evaporite minerals can be found in
the geological record back to 3.5~Ga, and since many of these were
originally precipitated in the form of gypsum (CaSO$_4\cdot2$H$_2$O),
which is converted to anhydrite (CaSO$_4$) at temperatures above
$58^\circ$C in pure water (and at lower temperatures in seawater),
temperatures cannot have been higher than this value
\citep{Holland1978, Holland1984, Walker1982}. Secondly, the continued
presence of life and the typical heat tolerance of living organisms
can be used to estimate an upper limit in the range of $40-60^ \circ$C
\citep{Walker1982}.

In conflict with these upper limits from evaporites and the continued
presence of life, low values of the $\delta^{18}$O isotope ratio in
3.5 to 3.0~Ga cherts were interpreted by some researchers as evidence
of a hot climate with oceanic temperature of $55-85^\circ$C
\citep{Knauth1976, Karhu1986, Knauth2003, Robert2006}. There is a lot
of debate, however, about how strongly oxygen isotope ratios actually
constrain temperatures. It has been argued, for example, that these
data could reflect a low $\delta^{18}$O of ancient seawater rather
than a hot climate \citep[see e.g.,][for discussions]{Walker1982,
  Kasting2006a, Kasting2006b, Kasting2006c, Jaffres2007}. An
alternative explanation for changes in isotope ratios during the
Precambrian has been put forward by \citet{vandenBoorn2007} who argue
that the data might reflect more widespread hydrothermal activity on
the ancient seafloor.

In light of this discussion there appears to be no strong argument in
favor of a hot Archean climate. Indeed, a recent analysis combining
oxygen and hydrogen isotope ratios indicates ocean temperatures below
40$^\circ$C for a sample of 3.4~Ga old rock
\citep{Hren2009}. \citet{Blake2010} analyzed $\delta^{18}$O isotope
compositions of phosphates in $3.2-3.5$~Gyr-old sediments and
interpreted the high $\delta^{18}$O found in their samples as being
indicative of low oceanic temperatures in the range
$26-35^\circ$C. These temperatures are close to the maximum of the
annually averaged sea-surface temperature of about 30$^\circ$C today
\citep{WOA2009_Temp}.

Although the evidence appears to point towards a temperate Archean
climate, the question of how warm the early Earth's atmosphere was is
certainly not quite settled yet. One further major problem is that
oceanic temperatures are expected to strongly vary with latitude and
depth. It is unknown, however, at what latitudes and depths the rocks
were formed on which the temperature estimates discussed above are
based. Notwithstanding these problems, the Archean climate was almost
certainly warm enough to keep the ocean surface from freezing
completely despite the low solar luminosity.

\subsection{Why was the Early Earth not Frozen?}

\citet{Donn1965} were, to my knowledge, the first to point out the
apparent discrepancy between the low solar luminosity predicted for
the young Sun and the evidence for liquid water on early Earth. Not
believing in a strong greenhouse effect in the early atmosphere, they
speculated that it could be used to put constraints on solar models
and theories of continental formation, an idea that certainly appears
rather optimistic from today's perspective.

As mentioned above, after the development of simple energy-balance
climate models by the end of the 1960s scientists began to study the
connection between a slight decrease in solar luminosity and
glaciations on Earth \citep[e.g.,][]{Budyko1969, Sellers1969}, but the
results had been discussed in the context of the Quaternary glaciation
rather than the climate of early Earth. The predictions of solar
models for a faint young Sun went not unnoticed in the planetary
science community, however: \citet{Pollack1971} investigated the
effect of the lower solar luminosity on the early atmosphere of Venus.

\begin{figure*}
\centerline{\includegraphics[width=10cm]{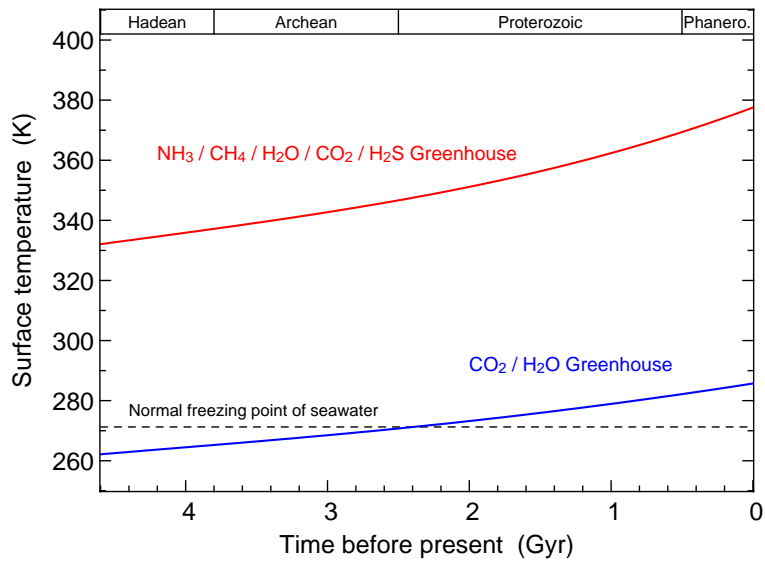}}
\caption{Average surface temperature evolution of Earth as a function
  of time given the changes in solar luminosity and assuming
  present-day concentrations of carbon dioxide and water vapor
  \textit{(blue line)} according to \citet{Sagan1972}. The
  calculations follow equation~(\ref{e:ebudget}) and assume a total
  pressure of 1~bar, an atmospheric composition constant with time and
  a fixed albedo of 0.35. Going into the past, the surface temperature
  drops below the normal freezing point of water $\sim 2$~Ga in this
  model. The solution of the faint young Sun problem suggested by
  \citet{Sagan1972} in terms of greenhouse gas warming dominated by
  ammonia (\amm, see Section~\ref{s:ammonia}) is shown as well
  \textit{(red line)}. In this scenario, volume mixing ratios
  $10^{-5}$ of \amm, \methane\ and \htwos\ have been added to the
  \cotwo-\water\ greenhouse. In terms of warming, ammonia is the
  dominant greenhouse gas in this case, Modified after
  \citet{Sagan1972}.}
\label{f:archeannh3}
\end{figure*}

One year later, \citet{Sagan1972} explored the effects of the lower
luminosity of the young Sun on the early Earth. \citeauthor{Sagan1972}
are usually credited as having discovered the faint young Sun
problem. While this is not entirely true as the discussion above
shows, they were certainly the first to make it known to a wider
public and to suggest a solution in terms of an enhanced greenhouse
effect. For their analysis, they used the fundamental equation for
Earth's global energy balance (see equation~(\ref{e:ebudget}) in
Section~\ref{s:solutions}), finding that global surface temperature
should have remained below the freezing point of sea-water for the
first two billion years of Earth's history with today's greenhouse gas
concentrations and albedo (see Figure~\ref{f:archeannh3}).  These
calculations neglected any feedback effects from water vapor and
included only simplified representations of the ice-albedo feedback
effect, however, so the problem was considered to be even more severe
in reality.

This conundrum of liquid water in a climate powered by a feeble Sun
has been termed the `faint young Sun problem' \citep{Ulrich1975},
sometimes also the `faint early Sun problem' or `faint young Sun
paradox'.  It is only a paradox, of course, if the Sun indeed was much
fainter in the early days of the solar system (alternative theories
are discussed in Section~\ref{s:massloss}), and if the parameters
controlling the climate in the Archean were similar to today's values,
an assumption which appears naive considering the profound changes
Earth has experienced in its long history. Indeed, changes in
atmospheric concentrations of greenhouse gases are one of the
possibilities to resolve the apparent inconsistency between the faint
young Sun and the temperate climate on early Earth (see
Section~\ref{s:greenhouse}).

Note that there is also evidence for the presence of liquid water
during several periods in the history of Mars, including at very early
times \citep{Carr1996}. The problem of keeping early Mars warm would
be even more profound due to its larger distance from the Sun and
considerably smaller mass, if there were indeed extended periods of
warm climate on early Mars. The faint young Sun problem for Mars,
however, will not be discussed in this review paper.

\section{Warming the Early Earth}
\label{s:solutions}

Before discussing possible solutions to the faint young Sun problem in
detail, it is helpful to ask which parameters govern the temperature
of Earth's atmosphere. The mean surface temperature $T_\mathrm{s}$ of
the atmosphere can be approximated by the following equation for the
case of a gray atmosphere, where the infrared absorption by greenhouse
gases is assumed to be independent of wavelength, \citep{Emden1913,
  Milne1922, Wildt1966, Stibbs1971}:

\begin{equation}
  \label{e:ebudget}
  T_\mathrm{s}^4 \: = \: \frac{1}{\varepsilon\sigma} \, \frac{R^2}{4} \,
  \frac{L_\odot}{4\pi r^2} \, (1-A)
  \, \left( \frac{}{} 1 + \frac{3}{4} \tau^\ast \frac{}{} \right)
\end{equation}

The variables in this equation are the Earth's effective surface
emissivity $\varepsilon$, the Stefan-Boltzmann constant $\sigma$, the
Earth's radius $R$, the solar luminosity $L_\odot$, the average
distance $r$ between Sun and Earth, the albedo $A$, and the column
infrared gray opacity $\tau^\ast$ representing the warming effect of
greenhouse gases.

Although solar energy is by far the most important source of energy
for today's climate system, other sources of energy (like heat from
Earth's interior, tidal energy from the gravitational interaction with
the Sun and the Moon, or the energy released after impacts from space)
could, in principle, provide additional heating to the early
atmosphere. Today, the globally integrated heat loss from the Earth's
interior (mostly originating from radioactive decay) amounts to less
than $5 \times 10^{13}$~W \citep{Pollack1993, Davies2010}. Taking into
account Earth's surface area of $5.1 \times 10^{14}$~m$^2$, this
corresponds to $\sim 0.1$~W~m$^{-2}$, more than three orders of
magnitude smaller than the climate forcing due to solar irradiance of
$\simeq 240$~W~m$^{-2}$ for the present-day solar constant (the solar
irradiance at the top of the atmosphere) of
$1361$~W~m$^{-2}$\citep{Kopp2011}. Total heat flow due to radioactive
decay is estimated to be factors of $\sim 3$ and $\sim 2$ higher than
today for the early and late Archean, respectively
\citep{Taylor2009}. Thus even on the early Earth, flows of internal
heat are at least two orders of magnitude too small to compensate for
the faint young Sun \citep{Endal1982}.

Dissipation of tidal energy amounts to about $3.5 \times 10^{12}$~W
(or $\sim 0.007$~W~m$^{-2}$) today \citep{Munk2007}, one order of
magnitude smaller than the geothermal heat flux. During the Archean,
tides were higher due to the smaller distance $r_\mathrm{M}$ to the
Moon \citep{Walker1986} which also influences the dissipation rate of
tidal energy due to the changes in orbital period and day-length
\citep{Zahnle1987}. The tidal energy dissipation rate during the
Archean can be estimated from the equations in \citet{Munk1968} and
the evolution of lunar distance \citep{Walker1986} to be a factor of
$\sim 3$ higher because of these effects, yielding an energy flux of
$\sim 0.02$~W~m$^{-2}$, again insufficient to provide enough energy to
counteract the lower solar irradiation.

The energy deposited by impactors from space can be estimated by
integrating the impact probability distribution \citep{Stuart2004}
over all energies, yielding an insignificant contribution to the
energy budget of about $5 \times 10^8$~W (corresponding to only
$10^{-6}$~W~m$^{-2}$) for the recent geological history. Impacts were
much more frequent very early in Earth's history, but the frequency of
major impacts from space had decreased dramatically with the end of
the Late Heavy Bombardment ($\sim 4.0$ to $3.8$~Ga) already before the
beginning of the Archean \citep{Tera1974, Wetherill1975, Sleep1989,
  Hartmann2000, Kring2002}, and was at most one order of magnitude
higher than today after the Late Heavy Bombardment, too low to deliver
significant amounts of energy globally \citep{Hartmann2000,
  Valley2002}. Major impacts could result in occasional melting of
frozen oceans during the Archean, however \citep{Bada1994}. Given the
widespread evidence for liquid surface water during the Archean,
episodic melting appears to be an unsatisfactory solution to the faint
young Sun problem.

Values for the surface emissivity $\varepsilon$ are very close to one
and do not vary much between different surface types
\citep[e.g.,][]{Wilber1999}, so any variations in $\varepsilon$ between
four billion years ago and today are small and cannot contribute
significantly to the solution of the faint young Sun
problem. Furthermore, a long-term increase in Earth's orbital radius
$r$ since the Archean seems unlikely. While planetary migration
through exchange of angular momentum is a widely discussed feature of
current models for the formation phase of planetary systems, it is
limited to the comparatively short formation time during which there
is a protoplanetary disk and planetesimals with which an exchange of
angular momentum is possible \citep{Papaloizou2006}.

Extremely speculative hypotheses like a potential variation of the
gravitational constant with time avoiding a faint young Sun
\citep{Newman1977, Tomaschitz2005} appear unlikely and will not be
discussed here.

In summary, a solution to the faint young Sun problem requires a
higher solar luminosity $L_\odot$ than predicted by standard solar
models, a lower overall albedo $A$ (and therefore increased absorption
of solar radiation) or a significantly enhanced greenhouse effect,
i.e., increased infrared opacity $\tau^\ast$ (or a combination of
these). All these have been suggested in the literature and will be
discussed in the remainder of this review article.

\section{Modifications of the Standard Solar Model}
\label{s:massloss}

The faint young Sun problem originates from the fact that the standard
solar model implies a considerably lower luminosity for the early
Sun. If the solar luminosity were higher than predicted by the
standard solar model, however, there might be no problem at all
\citep{Ulrich1975}.

The steady increase in solar luminosity with time shown in
Figure~\ref{f:sollum} is a fundamental corollary of the physical
equations governing the structure of and energy conversion within
stars (see Section~\ref{s:sun}). The only escape route
appears to be a change in stellar mass, since the luminosity $L$ of a
star powered by nuclear fusion of hydrogen to helium in its core (so
called main-sequence stars) steeply increases with its mass $M$
according to

\begin{equation}
\label{e:masslum}
  L \propto M^\eta \; , \mbox{where} \: \eta \simeq 2-4
\end{equation}

\citep{Kippenhahn1994}. $\eta$ depends on the mass of the star; for
stars like the Sun, a value of $\eta \simeq 4$ is usually
adopted. According to this mass-luminosity relation, a higher mass of
the young Sun would therefore go hand in hand with a higher initial
solar luminosity and would have the potential to avoid the faint young
Sun problem. Indeed, a higher initial mass together with an enhanced
mass loss of the early Sun has been suggested to avoid the faint young
Sun problem \citep{Boothroyd1991, Graedel1991}.

The present-day Sun loses mass due to two processes. First, hydrogen
is converted to helium in its core. The mass of the resulting helium
nucleus is less than the total mass of the protons entering this
fusion reaction, and the energy difference corresponding to this mass
difference is emitted by the Sun. Secondly, mass is continuously
transported away from the Sun by the solar wind, a stream of charged
particles (primarily electrons and protons) originating in the Sun's
upper atmosphere.

The mass-loss rates due to nuclear fusion and the solar wind amount to
$\dot{M}_\mathrm{fusion} \simeq 7 \times 10^{-14} \, M_\odot \,
\mathrm{yr}^{-1}$ and $\dot{M}_\mathrm{wind} \simeq 2 \times 10^{-14}
\, M_\odot \, \mathrm{yr}^{-1}$ \citep{Wood2004}, respectively,
yielding a total mass loss of $\dot{M} \simeq 1 \times 10^{-13} \,
M_\odot \, \mathrm{yr}^{-1}$ for today's Sun ($M_\odot\simeq 2 \times
10^{30}$~kg denotes the present-day solar mass). Assuming that this
mass-loss rate has not changed over the main-sequence lifetime of the
Sun, this would result in a solar mass only 0.05\% higher 4.57~Ga,
yielding a negligible increase in luminosity according to
equation~(\ref{e:masslum}).

The solar wind, however, is known to have been stronger for the young
Sun because of the higher solar activity in the past (see
Section~\ref{s:solutions}). Depending on the assumed mass-loss
history, a young Sun with an initial mass $\sim 4$\% higher than today
would be bright enough to explain the presence of liquid water on Mars
3.8~Ga \citep{Sackmann2003}, and an initial mass of $\sim 6$\% higher
than today makes the Sun as bright as today 4.5~Ga, although the solar
luminosity would still drop below today's levels during the Archean
\citep{Guzik1987, Sackmann2003}.  In addition to the direct increase
in energy input due to the higher solar luminosity, Earth would also
be closer to a more massive Sun on its elliptical orbit, further
enhancing the warming effect, with the semi-major axis $a \, (t)$ at
time $t$ inversely proportional to the solar mass $M \, (t)$
\citep{Whitmire1995}

\begin{equation}
\label{e:semimajor}
  a \, (t) \; \propto \; \frac{1}{M \, (t)} .
\end{equation}

There are limits to the mass of the early Sun, however. A weak upper
limit can be derived from the fact that at higher solar luminosities
Earth would have run into a runaway greenhouse effect \citep[see][for
a recent review]{Goldblatt2012}. If the solar luminosity were beyond a
certain threshold, the increased evaporation of water would result in
accelerating warming. Eventually, all ocean water would be evaporated
and lost to space by photodissociation and hydrodynamic escape, a
process which is believed to be responsible for the lack of water in
the atmosphere of Venus \citep{Ingersoll1969, Rasool1970}.

It has been estimated that a 10\% increase in solar flux could have
led to rapid loss of water from the early Earth
\citep{Kasting1988}. Taking into account the mass-luminosity relation
in equation~(\ref{e:masslum}), the change in Earth's semi-major axis
due to solar mass change from equation~(\ref{e:semimajor}) and the
secular evolution of solar luminosity following
equation~(\ref{e:sollum}), this corresponds to a 7\% increase in solar
mass \citep{Whitmire1995}, so high mass loss could make the Archean
unsuitable for life.

Furthermore, it has been suggested \citep{Guzik1995} that an extended
mass loss of the early Sun can be ruled out using helioseismology, the
study of the Sun's interior structure using resonant oscillations
\citep{Deubner1984}. Solving the faint young Sun problem would require
that the Sun remained at least a few percent more massive than today
over one or two billion years, while helioseismology limits the
enhanced mass loss to the first 0.2~Gyr of the Sun's life
\citep{Guzik1995}. A more extended period of mass loss leads to
changes in the distribution of heavier elements below the solar
convection zone, resulting in differences between calculated and
observed oscillation frequencies. \citeauthor{Guzik1995}'s model of
the interior of the Sun has been criticized by \citet{Sackmann2003},
however, who claim that models with initial masses up to 7\% higher
than today are compatible with helioseismological observations.

Much more stringent limits to a more massive young Sun can be inferred
from observations of mass loss in young stars similar to the Sun
\citep{Wood2004, Wood2005}. Observations of other cool stars show that
they lose most of their mass during the first 0.1~Gyr
\citep{Minton2007}. Most importantly, the observed solar analogs
exhibit considerably lower cumulative mass-loss rates than required to
offset the low luminosity of the early Sun \citep{Minton2007}. The
solution to the faint young Sun problem therefore seems to lie in the
other parameters controlling Earth's surface temperature, for example
the concentration of greenhouse gases in the early atmosphere, rather
than in a modification of the standard solar model involving higher
mass-loss rates.

\section{Enhanced Greenhouse Effect}
\label{s:greenhouse}

In today's climate, the temperature of Earth's troposphere is
increased due to the absorption of long-wave radiation from the
surface by atmospheric gases like water vapor, carbon dioxide, and
methane. This greenhouse effect \citep{Mitchell1989} has a natural and
an anthropogenic component. The natural greenhouse effect is the cause
for global average temperatures above the freezing point of water over
much of the Earth's history, while the anthropogenic component
resulting from the continuing emission of greenhouse gases by humanity
is responsible for the observed global warming since the 19th century
\citep{IPCC-4-1}.

Therefore, one obvious possibility to explain a warm early atmosphere
despite a lower insolation is an enhanced warming effect due to
atmospheric greenhouse gases like ammonia (\amm), methane (\methane),
or carbon dioxide (\cotwo).

\subsection{Ammonia}
\label{s:ammonia}

Ammonia is a very powerful natural greenhouse gas \citep{Wang1976}
because it has a strong and broad absorption feature at $\sim
10$~$\mu$m coincident with the peak in black-body emission from
Earth's surface. Ammonia seemed an attractive solution to the faint
young Sun problem in early studies for a number of historic
reasons. Indeed, in their original paper on the faint young Sun
problem, \citet{Sagan1972} suggest that an ammonia greenhouse could
have compensated the lower solar irradiance to keep Earth's oceans
from freezing over.

Historically, the choice of greenhouse gases like \amm\ (and \methane\
discussed in Section~\ref{s:methane}) as greenhouse gases was
motivated by three arguments: the assumption that the early atmosphere
was reducing, the apparent requirement of a reducing atmosphere for
the production of organic molecules, and the widespread glaciations at
the beginning of the Proterozoic. These historic arguments will be
explored in the following.

The view held at that time that Earth's early atmosphere was reducing
is closely linked to theories of planetary formation. Earth was formed
by accretion of smaller bodies (planetesimals) formed in the solar
nebula \citep{Wetherill1990} and may have formed a primary atmosphere
from gases (predominantly hydrogen) present in the nebula. This
primary atmosphere (if present) was quickly lost, however, and the
secondary atmosphere was generated by outgassing of volatiles
originally contained as chemical compounds within the planetesimals
\citep{Kasting1993}. A few decades ago, the accretion was believed to
have been slow, leading to a late formation of Earth's iron core. The
iron would thus have remained in the mantle for some time and favored
the formation of reducing gases that could then have accumulated in
the early atmosphere.

It is now believed, however, that the early atmosphere was not
strongly reducing. It was already pointed out in the 1960s and 1970s
that geochemical evidence for such an atmosphere is lacking
\citep{Abelson1966} and that a strongly reducing atmosphere of the
early Earth is unlikely, since the geochemistry of the upper mantle
and the crust suggest that the material was not in contact with
metallic iron \citep{Walker1976}. This implies a rapid formation of
Earth's iron core and an oxidation state of the mantle and the
atmosphere not too different from today. An additional argument in
favor of a fast accretion of Earth is the early formation of the Moon
$\sim 4.5$~Ga \citep{Canup2004}. The prevailing theory for the
formation of the Moon suggests that the Moon was created when a large
(roughly Mars-sized) impactor hit the young Earth \citep{Hartmann1975,
  Cameron1976, Canup2004}, requiring the accumulation of a significant
amount of material before the impact.

The second historically common argument in favor of a reducing early
atmosphere is that reducing gases appeared to be required for the
formation of the building blocks of life through lightning
\citep{Orgel1998, Chyba2010}. The famous experimental demonstration
that electric discharges in a strongly reducing gas mixture containing
methane, ammonia and hydrogen (then believed to resemble the early
atmosphere) produce a variety of simple organic molecules
\citep{Miller1953, Miller1959} led many to believe in this scenario
for the origin of prebiotic molecules. Indeed, in early papers on the
faint young Sun problem \citet{Sagan1972} remark that ammonia is ``a
very useful precursor compound for prebiological organic chemistry''
and \citet{Sagan1977} states that ``reduced atmospheric components
such as \amm\ and \methane\ are required to understand the
accumulation of prebiological organic compounds necessary for the
origin of life''.

There are other scenarios for the production of prebiotic molecules
which present viable alternatives to the Miller--Urey pathways,
however. One possibility is that organic molecules were delivered by
meteorites (in particular carbonaceous chondrites) or synthesized
during impacts \citep{Chyba1992}. Another scenario for the production
of biological precursor molecules relies on prebiotic chemistry taking
place in deep-sea hydrothermal vents, arguably the most likely
location for the origin of life anyway \citep{Martin2008}. Finally it
should be noted that substantial amounts of organic compounds like
formaldehyde (CH$_2$O) \citep{Pinto1980} and hydrogen cyanide (HCN)
\citep{Abelson1966, Zahnle1986} are photocemically produced even in
weakly reducing atmospheres, where the latter requires the presence of
methane. It should be noted that a very low atmospheric ammonia
concentration of $\sim 10^{-8}$ required for the evolution of life
\citep{Bada1968} can also be maintained in an atmosphere with high
concentrations in carbon dioxide \citep{Wigley1981}.

The third frequently used argument in favor of reducing greenhouse
gases like \amm\ and \methane\ is that the major ``Huronian''
glaciations of the planet occurring in the time interval $2.4-2.2$~Ga
could have been triggered by the first major rise of atmospheric
oxygen \citep[see][for recent reviews]{Canfield2005, Catling2005,
  Holland2006, Sessions2009} around the same time. The increase in
atmospheric \otwo\ would have dramatically diminished the
concentration of \methane\ and other reducing greenhouse gases like
\amm\ via oxidation, resulting in global cooling \citep{Kasting1983,
  Pavlov2000, Kasting2001, Kasting2005, Kopp2005, Haqq-Misra2008}. The
relative timing of these events is obviously crucial, and it now
appears that the first global glaciation occurred close to 2.4~Ga
\citep{Kirschvink2000} and thus \textit{before} the Great Oxidation
Event which is dated closer to 2.3~Ga and thus around the time of the
second of three Huronian glaciations \citep{Bekker2004}. Furthermore,
there is evidence for an even earlier continental glaciation from
glacial deposits in the Pongola Supergroup dated $\sim 2.9$~Ga
\citep{Young1998} although it remains unclear whether this was a
global event. Distinct pulses of oxygenation associated with those
glaciations might explain these findings; in any case, the argument is
not as clear-cut as often suggested.

Thus most of these historic arguments in favor of ammonia (and other
reducing greenhouse gases) have now been put into perspective, but it
remains interesting to see how much ammonia would be required to
offset the faint young Sun. In their paper, \citet{Sagan1972} used a
simple two-layer approximation to the atmosphere's energy budget to
show that early Earth could have been kept warm by very low partial
pressures ($p_\mathrm{\,NH_3} = 10^{-5}$~bar) of ammonia added to an
atmosphere with a total pressure of 1~bar and today's concentrations
of carbon dioxide and water vapor (\water) as well as volume mixing
ratios of $10^{-5}$ of methane and hydrogen sulfide (\htwos), see
Figure~\ref{f:archeannh3}. For comparison, the partial pressure of
ammonia in the present-day atmosphere is only $6 \times 10^{-9}$~bar
\citep{Wang1976}. The Archean Earth surface temperatures of $T_s
\simeq 340$~K derived in \citet{Sagan1972} for their ammonia-dominated
greenhouse are actually considerably above the normal freezing point
of water.

Despite its strong warming effect, subsequent studies of the faint
young Sun problem revealed difficulties with ammonia as the dominant
greenhouse gas in the Archean. \citet{Kuhn1979} used a more
sophisticated radiative transfer model and confirmed the results of
\citet{Sagan1972} by showing that \amm\ partial pressures larger than
$p_\mathrm{\,NH_3} = 8 \times 10^{-6}$~bar for an albedo of 0.30 and
an atmosphere with a total pressure of 0.78~bar, present-day water
vapor content and a carbon dioxide partial pressure of
$p_\mathrm{\,CO_2} = 3.6 \times 10^{-4}$~bar are sufficient to keep
Earth from freezing. They pointed out one significant problem,
however, which had earlier been noted by \citet{Abelson1966}: using
models for the photochemistry of ammonia, they demonstrated that the
Sun's ultraviolet radiation (which was much more intense during the
Archean, see Section~\ref{s:sun}) would have destroyed this amount of
\amm\ via photodissociation in less than a decade. They conclude that
continuous outgassing of ammonia from the Earth's interior would have
been required to make an \amm\ greenhouse during the Archean
work.

Investigating this balance between outgassing and photochemical
destruction, \citet{Kasting1982} estimated steady-state ammonia
formation rates for the early Earth and concluded that abiotic sources
could have been sufficient to sustain mixing ratios of $\sim 10^{-8}$
which have been argued to be required for the evolution of life in the
ocean based on the rapid decomposition of aspartic acid in the absence
of ammonium and the assumption that aspartic acid is necessary for
life to originate \citep{Bada1968}. The ammonium resupply rates
derived in \citet{Kasting1982} are insufficient to provide substantial
greenhouse warming, however.

It should also be noted that ammonia is highly soluble
\citep{Levine1980} and thus quickly rained out of the atmosphere and
dissolved as ammonium (NH$_4^+$) in the oceans \citep{Kasting1982,
  Walker1982}. Sustaining atmospheric partial pressures of ammonia in
the range required to offset the faint young Sun requires $0.1-10$
percent of the atmospheric nitrogen to be dissolved in the ocean [C.\
Goldblatt, private communication].

Due to these problems, ammonia had fallen out of favor as the dominant
greenhouse gas in the Archean atmosphere. More recently,
\citet{Sagan1997} revived the idea of an Archean ammonia greenhouse by
pointing out that an early atmosphere containing nitrogen (\ntwo) and
\methane\ would form an organic haze layer produced by
photolysis. This layer would block ultraviolet radiation and thus
protect \amm\ from photodissociation. Others showed, however, that the
existence of such a layer would lead to an `anti-greenhouse' effect
because it blocks solar radiation from reaching the surface but allows
thermal radiation to escape to space \citep{McKay1991,
  McKay1999}. High humidity has been shown in experimental studies to
further enhance this cooling effect of aerosols
\citep{Hasenkopf2011}. Furthermore, the size distribution of the haze
particles could have limited the layer's shielding function against
solar ultraviolet radiation \citep{Pavlov2001}, although laboratory
experiments suggest particle sizes which make the haze optically thick
in the ultraviolet yet optically thin in the optical
\citep{Trainer2006}.

The ammonia story took an unexpected turn recently, when
\citet{Ueno2009} suggested that carbonyl sulfide (OCS) at ppmv (parts
per million volume) levels could explain the distribution of sulfur
isotopes in geological samples from the Archean and could shield \amm\
against ultraviolet radiation. Detailed photochemical modeling shows,
however, that such high concentrations of OCS are unlikely because OCS
is rapidly photodissociated in the absence of ultraviolet shielding by
ozone \citep{Domagal-Goldman2011}.

As an additional argument against the cooling effects of haze layers,
\citet{Wolf2010} demonstrated in a general circulation model with
size-resolved aerosols that the fractal structure of the aerosol
particles forming the haze drastically diminishes the anti-greenhouse
effect. Such fractal particles give a good fit to the albedo spectrum
of Titan, the largest moon of Saturn, which has a dense atmosphere
with an opaque organic haze layer \citep{Danielson1973, Rages1980,
  Rages1983, McKay1991}. In addition, \citet{Hasenkopf2011} showed
that the aerosol particles in the haze could have led to the formation
of short-lived and optically thin clouds with a lower albedo than
today's clouds, hence decreasing their cooling and increasing their
warming effect. (Note, however, that cloud effects alone are
insufficient to effectively counteract the faint young Sun, see
Section~\ref{s:clouds}.)

In summary, ammonia may not be completely out of the game as a
possible solution of the faint young Sun problem after all, although
potential problems with the haze shielding and the high solubility of
ammonia appear to make \methane\ and \cotwo\ more likely candidates.

\subsection{Methane}
\label{s:methane}

Given the problems with ammonia as a greenhouse gas in the Archean,
some researchers turned to methane (\methane) as a potential warming
agent for the Archean climate.

\begin{figure*}[t]
\centerline{\includegraphics[width=12cm]{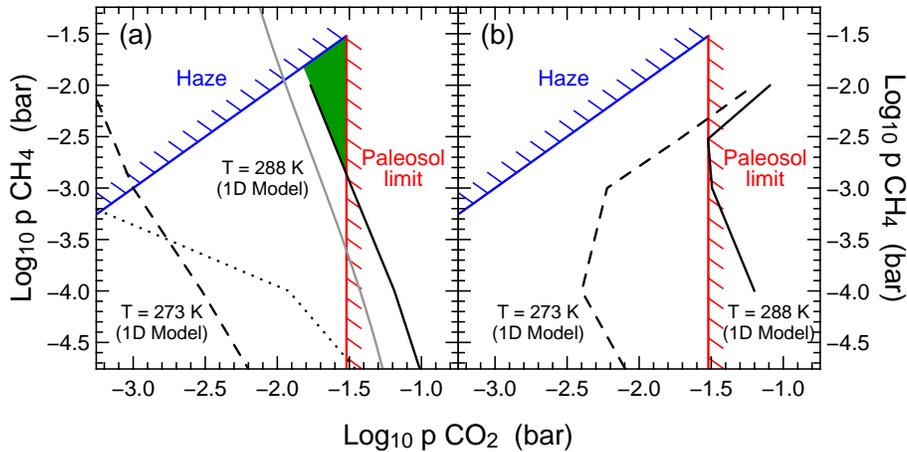}}
\caption{Parameter space for a late-Archean methane
  greenhouse. \textbf{(a)} The \textit{solid black line} shows the
  methane and carbon dioxide partial pressures required to sustain a
  global average surface temperature $T = 288$~K based on the model
  calculations for a \methane--\cotwo--\water\ greenhouse at 1~bar
  total pressure and a solar luminosity of $0.8 \, L_\odot$
  corresponding to a time 2.8~Ga \citep{Haqq-Misra2008}; the
  \textit{dashed line} indicates the partial pressures for $T =
  273$~K. Earlier model calculations for $T = 288$~K are also shown:
  the results from \citet{Kiehl1987} as \textit{gray line} and the
  erroneous results from \citet{Pavlov2000} as \textit{dotted
    line}. The limit of haze formation is indicated in \textit{blue},
  the paleosol upper limit on the \cotwo\ partial pressure in
  \textit{red}. The small \textit{green triangle} shows the possible
  parameter space for greenhouse warming sufficient to prevent global
  glaciation given the constraints from haze formation and paleosol
  geochemistry. \textbf{(b)} Same as (a), but for model calculations
  explicitly taking into account additional warming by ethane
  (C$_2$H$_6$) and cooling by organic haze \citep{Haqq-Misra2008}. A
  total atmospheric pressure $p = 1$~bar is assumed when converting
  from volume mixing ratios to partial pressures.}
\label{f:archeanch4}
\end{figure*}

The main advantage of methane as compared to ammonia discussed in
Section~\ref{s:ammonia} above is that \methane\ is photolyzed
considerably slower than \amm, because it requires ultraviolet light
of much shorter wavelengths ($\lesssim 145$~nm) where the Sun emits
less radiation. Indeed, photochemical models show that even under the
more intense ultraviolet radiation emitted by the young Sun, the
lifetime of \methane\ in a terrestrial atmosphere low in \cotwo\ is of
the order of $10^3$ to $10^4$~years \citep{Zahnle1986}, in contrast to
less than 10~years for \amm.

There are two effects constraining the allowed parameter space for a
methane greenhouse on the early Earth, though. First, depending on the
assumed atmospheric methane partial pressure, a contribution from
other greenhouse gases to the warming will be required, with carbon
dioxide being the most natural choice. As discussed in
Section~\ref{s:carbondioxide} below, geochemical data from ancient
paleosols set an upper limit to the atmospheric carbon dioxide partial
pressure of at most $p_\mathrm{\,CO_2} < 0.03$~bar during the late
Archean. Scenarios with low methane partial pressures could be in
conflict with this constraint, unless other forcings contribute to
warming.

Secondly, photochemical models show that an organic haze starts to
form at high \methane/\cotwo\ ratios \citep{Kasting1983}. As discussed
above, an organic haze layer exhibits an anti-greenhouse effect
because it reflects solar radiation back into space while being
transparent to outgoing infrared radiation \citep{McKay1991,
  McKay1999}. This haze would thus cool the planet, effectively
limiting the greenhouse warming achievable by methane in the early
atmosphere. Earlier photochemical modeling indicated that organic haze
should form in the primitive atmosphere at \methane/\cotwo\ ratios
larger than $1$ \citep{Zahnle1986, Pavlov2001}. Recent laboratory
experiments \citep{Trainer2004, Trainer2006} suggest that haze could
start to form at even lower mixing ratios of $\sim 0.2-0.3$. Note,
however, that the fractal nature of haze particles already discussed
in the context of a possible shielding of ammonia from ultraviolet
radiation would have limited the anti-greenhouse effect of the haze
layer \citep{Wolf2010}.

Moreover, laboratory experiments show that enhanced concentrations of
up to 15\% of hydrogen (H$_2$) decrease the amount of haze formed in a
\cotwo-rich atmosphere and thus limit the anti-greenhouse effect while
providing sufficient warming for the Archean Earth
\citep{DeWitt2009}. The amount of hydrogen in the early atmosphere is
determined by the balance between volcanic outgassing and hydrogen
escape to space. Conventional wisdom suggests that hydrogen escape on
early Earth is limited by upward diffusion \citep{Hunten1973,
  Walker1977}, resulting in atmospheric hydrogen mixing ratios of the
order of $10^{-3}$. It has been argued, however, that in the anoxic
early atmosphere temperatures at the base of exosphere (the outermost
atmospheric layer) would have been much lower, resulting in
considerably slower hydrogen escape and thus larger hydrogen mixing
ratios \citep{Watson1981, Tian2005}. \citet{Tian2005} estimated
molecular hydrogen mixing ratios of up to 30\% in the early
atmosphere. This notion of a hydrogen-rich early atmosphere remains
controversial, however \citep{Catling2006, Tian2006} and should be
investigated with photochemistry models which are more appropriate
than the models used so far. It is also unclear whether such a large
hydrogen inventory would be maintained in the presence of methanogenic
bacteria which consume hydrogen in their metabolism.

But how much methane would be required to warm the Archean atmosphere,
and how does this compare to the constraints from paleosols and haze
formation?  \citet{Kiehl1987} were the first to calculate the
potential contribution of methane to an Archean \cotwo\ greenhouse
(see Figure~\ref{f:archeanch4}a). According to their calculations,
\cotwo\ partial pressures of $p_\mathrm{\,CO_2} \sim 0.1$~bar and
$p_\mathrm{\,CO_2} \sim 0.03$~bar would be sufficient to reach average
surface temperatures similar to today for the early and late Archean,
respectively, when methane at a mixing ratio of $10^{-4}$ is present
in the atmosphere. These values for the carbon dioxide partial
pressure are about a factor of $\sim 3$ lower than without methane,
see the discussion in Section~\ref{s:carbondioxide}.

Quite a bit of confusion has been caused by the subsequent study by
\citet{Pavlov2000} which reported considerably stronger warming in a
late-Archean methane greenhouse as compared to \citet{Kiehl1987}, in
particular at higher methane partial pressures (see
Figure~\ref{f:archeanch4}a). Unfortunately, these results were due to
an error in the radiative-transfer code, and revised calculations
\citep{Haqq-Misra2008} show a methane warming that is actually smaller
(at a given methane concentration) than the earlier calculations by
\citet{Kiehl1987}. These model calculations are compared to the
constraints from haze formation and geochemistry of paleosols in
Figure~\ref{f:archeanch4}a, leaving only a small triangle in the $\log
p_\mathrm{CH_4}-\log p_\mathrm{CO_2}$ parameter space where sufficient
warming can be provided without cooling by organic haze and without
conflict with the paleosol constraints on $p_\mathrm{CO_2}$. Note
that, depending on temperature, the upper limits on carbon dioxide
partial pressure could be even lower (see Figure~\ref{f:archeanco2}
and the discussion below). It is less clear how tight the constraint
from haze formation is in reality as, on the one hand, haze could be
formed at even lower \methane/\cotwo\ mixing ratios
\citep{Trainer2004, Trainer2006}, but could exhibit a decreased
anti-greenhouse effect due to the fractal nature of the aerosol
particles forming the haze layer \citep{Wolf2010}, see the discussion
in Section~\ref{s:ammonia} above.

The recent model calculations by \citet{Haqq-Misra2008} taking into
account the anti-greenhouse effect of (non-fractal) organic haze
(which starts to form at \methane/\cotwo\ mixing ratios of $\sim 0.1$
in their model, in agreement with the laboratory results discussed
above) and additional warming by ethane (C$_2$H$_6$) are shown in
Figure~\ref{f:archeanch4}b. According to these simulations, a
late-Archean \cotwo--\methane\ solution to the faint young Sun problem
appears to be more complicated than previously thought because organic
haze formation sets in at higher methane partial pressures while high
carbon dioxide partial pressures are ruled out by paleosol
constraints, yielding insufficient warming to explain the absence of
glaciation in the late Archean. This strongly depends on the still
somewhat obscure properties of organic haze layers in the early
atmosphere, however, and other gases besides \cotwo\ and \methane\
might have contributed to the warming.

\begin{figure*}[t]
\centerline{\includegraphics[width=10cm]{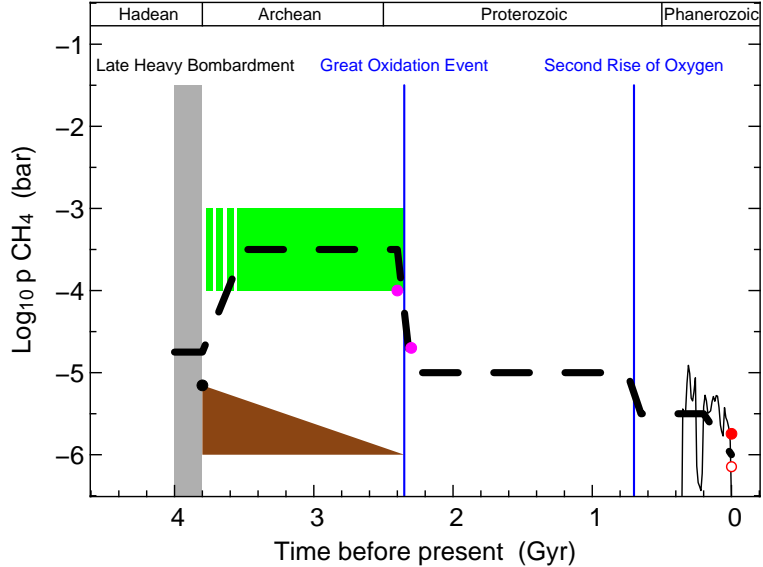}}
\caption{Estimates for the methane partial pressure in the atmosphere
  in various epochs in Earth's history. The period of frequent impacts
  during the Late Heavy Bombardment is shown in \textit{gray}, with
  the estimate for methane produced in impacts by \citet{Kasting2005}
  as \textit{black circle}. The \textit{green area} indicates the
  range based on estimates of biological methane fluxes during the
  Archean \citep{Kasting2005}. The \textit{brown triangle} shows the
  contribution from abiogenic sources based on the present-day
  estimate of \citet{Emmanuel2007}, including a possible increase up
  to a factor of 10 in earlier times due to faster creation of
  seafloor \citep{Kharecha2005}. The decrease with time is not based on
  any detailed model but only intended to give a rough indication of
  this possibility. Estimates for atmospheric methane content from a
  model of the Great Oxidation Event are indicated in \textit{magenta}
  \citep{Goldblatt2006}. Phanerozoic \methane\ concentrations
  estimated in \citet{Beerling2009} are represented by the
  \textit{thin black line}. Finally, the pre-industrial and
  present-day methane partial pressures are shown as \textit{open} and
  \textit{filled red circle}, respectively \citep{IPCC-4-1-2}. The
  \textit{thick black dashed line} is a highly idealized sketch of
  Earth's methane history based on these estimates. Methane fluxes
  have been converted to atmospheric mixing ratios using the relation
  shown in \citet{Kasting2005}, and a total pressure $p = 1$~bar is
  assumed for the conversion from volume mixing ratios to partial
  pressure values.}
\label{f:ch4history}
\end{figure*}

Finally, although methane is considerably more stable than ammonia, it
is continuously depleted by photolysis and reactions with hydroxyl
(OH) radicals. Thus it is interesting to ask what constraints on
Archean methane fluxes and atmospheric concentrations can be derived.

Before discussing estimates of atmospheric methane concentrations
during the Archean, we take a brief look at the methane budget of
today's atmosphere. The methane concentration in the present-day
atmosphere is about 1.8~ppmv, having increased from $\sim 0.7$~ppmv in
pre-industrial times due to anthropogenic methane emissions from
agriculture and industrial processes \citep{IPCC-4-1-2}. Methane
sources today amount to a methane flux of about 600~\tgyr\
($1\,\mathrm{Tg} = 10^{12}$~g) \citep{IPCC-4-1-7}. In the literature,
estimates for Archean methane fluxes are often compared to this
present-day flux (frequently and inaccurately even called the
``current biological flux''). This is of course a valid
order-of-magnitude comparison in principle, but it should be kept in
mind that more than 60\% of today's methane flux is from anthropogenic
sources (including industrial processes and emissions related to
fossil fuels), and about 90\% of the remaining natural flux originates
from ecosystems which were not present during the Archean, i.e.,
wetlands, termites, wild animals and wildfires \citep{IPCC-4-1-7}.

Today, methane is predominantly produced biologically. In the Archean,
three sources of methane have contributed to the atmospheric budget:
impacts from space, geological sources, and anaerobic ecosystems
\citep{Kasting2005}, see Figure~\ref{f:ch4history} for an overview.

Very high methane fluxes from cometary impacts of $\gtrsim 500$~\tgyr\
at 3.5~Ga and $\gtrsim 5000$~\tgyr\ at 3.8~Ga have been estimated by
\citet{Kress2004} based on impact rates derived in
\citet{Chyba1990}. More modest \methane\ production rates appear more
likely, however. \citet{Kasting2005} estimates an methane flux from
impacts at the beginning of the Archean 3.8~Ga of $\sim
20$~\tgyr. Using the non-linear relation between methane source flux
and atmospheric concentration based on photochemical modeling given
in \citet{Kasting2005} and based on \citep{Pavlov2001}, this
corresponds to a volume mixing ratio $\sim 7$~ppmv.

The order of magnitude of geological methane sources in the Archean
can be derived from the present-day abiogenic methane flux. The
current flux from mineral alteration at mid-ocean ridges, emissions
from volcanoes, and geothermal sources based on the most recent data
has been estimated to be $\sim 2.3$~\tgyr\ \citep{Emmanuel2007},
sufficient to sustain $\sim 1$~ppmv according to
\citet{Kasting2005}. In the early Archean, this flux could have been a
factor of 5 to 10 larger due to the faster creation of seafloor on
early Earth \citep{Kasting2005}, resulting in an atmospheric mixing
ratio of $\sim 7$~ppmv.

Therefore, low concentrations of the order of 10~ppmv of methane in
the atmosphere could have been sustained from abiogenic sources in the
early Archean. Later in time, after the origin of life and before the
first major rise in atmospheric oxygen, much larger methane
concentrations can be achieved from biological sources. Biological
methane production today is accomplished by methanogenic bacteria (or
methanogens for short) which are believed to have arisen very early in
the evolution of life \citep{Woese1977}. Their metabolism is based on
a variety of metabolic pathways \citep{Thauer1998}. The two most
important net reactions are

\begin{eqnarray}
  \nonumber
  \mathrm{CO}_2 \: + \: 4 \, \mathrm{H}_2 & \longrightarrow &
  \mathrm{CH}_4 \: + \: 2 \, \mathrm{H}_2\mathrm{O} \:\:\: \mathrm{and} \\
  \mathrm{CH}_3\mathrm{COO}^- \: + \: \mathrm{H}^+ & \longrightarrow &
  \mathrm{CH}_4 \: + \: \mathrm{CO}_2 .
\end{eqnarray}

Assuming that methanogens converted most of the hydrogen available in
the atmosphere \citep{Kral1998, Kasting2001} and using an estimated
hydrogen mixing ratio of $(1-2) \times 10^{-3}$, Archean methane
mixing ratios of $500-1000$~ppmv could be plausible. More elaborate
simulations with a coupled photochemistry-ecosystem model essentially
confirm these early estimates, with atmospheric methane mixing ratios
in the range $100-1000$~ppmv for reasonable atmospheric hydrogen
fractions \citep{Kharecha2005}. It should be noted that our
understanding of Archean ecosystems is naturally rather limited, so
these estimates should be taken with a grain of salt.

Nevertheless, from these arguments one can conclude that methane
mixing ratios in the Archean atmosphere of up to 1000~ppmv appear
plausible, see Figure~\ref{f:ch4history}. Comparing this to the
results from climate model simulations for the late-Archean presented
in Figure~\ref{f:archeanch4}, it is obvious that these are
insufficient to provide enough warming given the paleosol constraints
on carbon dioxide partial pressures during that time. Even if higher
methane fluxes should have been achieved, haze formation limits the
warming in a late-Archean methane greenhouse, although this depends on
the details of organic-haze formation and the properties of the
particles within the haze layer, see the discussion above. Note that
the production of haze is self-limiting, as more haze would cool the
climate and thus reduce the amount of methane produced by methanogens
\citep{Domagal-Goldman2008}.

In summary, it remains unclear whether methane could have provided
sufficient warming at least for the late Archean, but a solution of
the faint young Sun problem based on methane certainly appears to be
considerably more complicated than previously thought.

\subsection{Carbon Dioxide}
\label{s:carbondioxide}

Due to the increasing amounts of arguments against a strongly reducing
early atmosphere \citep{Walker1976}, carbon dioxide was suggested
early on as the dominant greenhouse gas counteracting the faint young
Sun on Earth \citep{Owen1979, Walker1981, Kuhn1983, Kasting1984a,
  Kasting1986, Kasting1987} and Mars \citep{Cess1980}. Carbon dioxide
is an attractive solution to the faint young Sun problem in the sense
that the long-term evolution of the atmospheric carbon dioxide
concentration is controlled by the inorganic carbon cycle, part of an
important negative feedback loop which stabilizes Earth's climate on
geological timescales \citep{Walker1981, Berner1983}.  The inorganic
carbon cycle removes \cotwo\ from the atmosphere via silicate
weathering according to the reaction

\begin{eqnarray}
  \nonumber
  \mathrm{CaSiO}_3 \; + \; 2 \, \mathrm{CO}_2 &+& \mathrm{H}_2\mathrm{O}
  \; \longrightarrow \\ \mathrm{Ca}^{++} &+& 2 \, \mathrm{HCO}_3^- \;
  + \; \mathrm{SiO}_2 .
\end{eqnarray}

(For illustrative purposes the silicate mineral wollastonite,
CaSiO$_3$, is taken here to represent all silicate rock.) The products
of this reaction are transported by rivers to the oceans, where they
are -- biotically or abiotically -- converted into calcium carbonate:

\begin{eqnarray}
  \nonumber
  \mathrm{Ca}^{++} + 2 \, \mathrm{HCO}_3^- \; \longrightarrow \;
  \mathrm{CaCO}_3 + \mathrm{CO}_2 + \mathrm{H}_2\mathrm{O} ,\\
\end{eqnarray}

resulting in the net formula for the so-called Urey
silicate-weathering reaction

\begin{equation}
\mathrm{CaSiO}_3 \; + \; \mathrm{CO}_2 \; \longrightarrow \;
\mathrm{CaCO}_3 \; + \; \mathrm{SiO}_2 . 
\end{equation}

This precipitated calcium carbonate is then partly deposited in
sediments at the bottom of the oceans. The sediments on the seafloor
are then transported via the motions of plate tectonics. At subduction
zones, most of the carbon dioxide is returned to the atmosphere via
arc volcanism, while some is incorporated into the Earth's mantle,
depending on the composition of the sediments and temperature
\citep{Kerrick2001, Stern2002}. Quite remarkably, the basic principles
of the inorganic carbon cycle were already discovered by several
scientists in the 19th century \citep[see][for discussions of this
early history of ideas about the inorganic carbon cycle]{Berner1995,
  Berner1996}.

The silicate-weathering cycle is part of a negative feedback loop
because the weathering rate removing \cotwo\ from the atmosphere
increases with growing atmospheric \cotwo\ concentrations and rising
temperatures (and vice versa), while the volcanic emission of \cotwo\
can be assumed to be roughly constant over geological time (when
averaged over sufficiently long timescales to suppress the large
variations caused by individual eruptions), or possibly decreasing
over time governed by changes in geothermal heat flow and volcanic
activity.

\begin{figure*}
\centerline{\includegraphics[width=10cm]{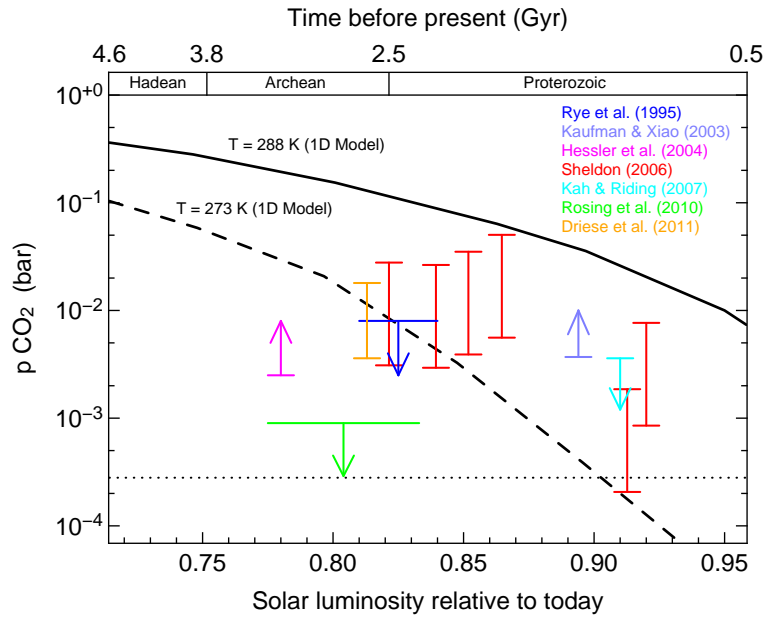}}
\caption{Comparison of empirical estimates of carbon dioxide partial
  pressures during the Precambrian and climate model results for an
  average global surface temperature of 288~K assumed to be required
  to prevent global glaciation as a function of relative solar
  luminosity \textit{(solid black line)}. The results for a global
  mean temperature of 273~K are indicated by the \textit{dashed black
    line}. Calculations are based on a one-dimensional
  radiative-convective climate model \citep{vonParis2008}. Geochemical
  estimates for atmospheric \cotwo\ partial pressures at different
  epochs are indicated \citep{Rye1995, Hessler2004, Sheldon2006,
    Rosing2010, Driese2011}, see the text for details. A temperature
  of 298~K is assumed in case an explicit dependence of the estimates
  on environmental temperature is available. In addition to the
  Archean and Paleoproterozoic estimates, four Mesoproterozoic
  estimates are shown for comparison: a lower limit derived from a
  carbon isotope analysis of microfossils dating back 1.4~Ga
  \citep{Kaufman2003}, a $\sim 1.2$~Ga upper limit inferred from
  in-vivo experiments of cyanobacterial calcification \citep{Kah2007}
  and two estimates from \citet{Sheldon2006}. The dotted line shows
  the pre-industrial \cotwo\ partial pressure of $2.8 \times
  10^{-4}$~bar. The conversion from solar luminosity (bottom scale) to
  age (top scale) follows the approximation given in
  equation~(\ref{e:sollum}). Modified and updated after
  \citet{Kasting2010}.}
\label{f:archeanco2}
\end{figure*}

Following the initial work on carbon dioxide in the Archean
atmosphere, one-dimensional radiative-convective climate models were
used to estimate the amount of \cotwo\ necessary to keep Earth from
freezing (see also Figure~\ref{f:archeanco2}). For solar luminosities
of $L = 0.75 \, L_\odot$ representative for the early Archean, these
models suggest that carbon dioxide partial pressures of
$p_\mathrm{\,CO_2} \simeq 0.3$~bar (or more than 1,000 times the
pre-industrial value of $p_\mathrm{\,CO_2} \simeq 0.00028$~bar) are
required to reach global average surface temperatures similar to
today, i.e., $T_\mathrm{s} \simeq 288$~K, whereas partial pressures of
$p_\mathrm{\,CO_2} \simeq 0.1$~bar (about 300 times the present-day
value) are sufficient for the late Archean \citep{Owen1979,
  Kasting1984a, Kiehl1987, vonParis2008}.

A temperature of 288~K would presumably correspond to a world with
small ice caps similar to our present climate, the limit of complete
freezing is often set at a mean surface temperature of 273~K, the
freezing point of water. Carbon dioxide values required to reach this
temperature are typically $p_\mathrm{\,CO_2} \simeq 0.06$~bar (or
about 200 times pre-industrial levels) for the early Archean and
$p_\mathrm{\,CO_2} \simeq 0.01$~bar (roughly 30 times pre-industrial
levels) for the late Archean, respectively.  These numbers are
generally interpreted as lower limits, since the ice-albedo feedback
(and many other factors) are not adequately considered in these
calculations, but note that some studies \citep[e.g.,][]{Kasting1987,
  Kasting1993} crudely account for the ice-albedo feedback effect by
requiring a minimum mean surface temperature of 278~K based on recent
glaciations, thus yielding higher \cotwo\ concentrations than the ones
reported for 273~K mean surface temperature reported above (0.1~bar
and 0.03~bar \cotwo\ partial pressure for the beginning and end of the
Archean, respectively).

Note that the inorganic carbon cycle operates on very long timescales,
so the question arises of whether such high carbon dioxide
concentrations are sufficient to \textit{stabilize} the climate in the
Archean. The timescales for increasing the atmospheric \cotwo\
concentration due to faster rates of volcanic outgassing and/or slower
rates of weathering ($\sim 10^5$~yr) are much longer than the ones for
the formation of snow and ice ($\sim 1$~yr), so any transient cooling
would lead to global glaciation \citep{Caldeira1992}. It has been
furthermore suggested that the formation of highly reflective \cotwo\
clouds in the atmosphere could make this glaciation irreversible.
Carbon dioxide ice clouds scatter solar radiation and thus raise the
albedo, but were assumed to be nearly transparent to thermal
radiation. They were therefore expected to cool the planet
\citep{Kasting1991}. This would be true if the clouds were composed of
particles smaller than a few micrometers in size, but larger particles
can be expected in such clouds which then would scatter infrared
radiation very effectively and thus result in a net warming effect
\citep{Forget1997}, provided that they are not low and optically thick
\citep{Mischna2000}.

Carbon dioxide as the dominant greenhouse gas offsetting the faint
early Sun has been criticized on two grounds, however. First, it has
been argued that the removal of atmospheric carbon dioxide during the
Archean was dominated by the flow of carbon into the mantle via the
subduction of carbonatized seafloor on a tectonically more active
Earth rather than silicate weathering \citep{Sleep2001}. This would
constantly diminish the atmospheric reservoir of \cotwo, thus
decreasing its warming effect on the Archean climate. This may not be
a major problem for the notion of carbon dioxide as main warming agent
during the Archean, however, since at low \cotwo\ levels (and in the
absence of other greenhouse gases) the flow of carbon dioxide from the
atmosphere to the ocean is limited by the ice cover on the oceans.

Secondly, and most importantly, geochemical analysis of paleosols and
banded iron formations provides constraints on the atmospheric \cotwo\
concentration during the late Archean to values much lower than
required to solve the faint young Sun problem, see
Figure~\ref{f:archeanco2}.  \citet{Rye1995} used the absence of
siderite (\siderite) and a thermodynamic model for the mineral
assemblage in 2.2--2.75 Gyr-old paleosols to establish an upper limit
of about 100 times present-day levels. During anoxic weathering of
basalt, iron is washed out of the upper layers of soils and either
transferred to the ground water or precipitated within the mineral
assemblage. At high \cotwo\ partial pressures, siderite would be
expected to be found in the lower parts, at low \cotwo\ levels the
iron would be precipitated in the form of iron silicates. Siderite is
absent from all of the paleosols older than 2.2~Gyr, however, whereas
other iron-rich minerals are found, suggesting low atmospheric \cotwo\
levels during the late Archean and early Proterozoic.

Conflicting evidence for a \cotwo-rich atmosphere during the Archean
and early Proterozoic based on the occurrence of siderite in banded
iron formations \citep{Ohmoto2004a, Ohmoto2004b} was convincingly
challenged \citep{Kasting2004, Sleep2004, Sheldon2006}.

Even lower values for the \cotwo\ levels in the early Proterozoic were
later derived by \citet{Sheldon2006} from an improved model relying on
the mass balance of weathering rather than the thermodynamic argument
used in \citet{Rye1995}, yielding a range of $p_\mathrm{\,CO_2} \sim
0.003 - 0.03$~bar for three samples of $\sim 2.2$~Gyr-old paleosols,
and values in a similar range for samples with ages from 2.5 to
1.8~Gyr. Applying the same method to a late-Archean paleoweathering
profile dated at 2.69~Ga yields a range of \cotwo\ partial pressures
$p_\mathrm{\,CO_2} \sim 0.004 - 0.02$~bar \citep{Driese2011},
consistent with the results obtained by \citet{Rye1995} and
\citet{Sheldon2006}.

An even lower upper limit for the \cotwo\ partial pressure in the
Archean atmosphere was derived by \citet{Rosing2010}. In this paper,
the authors argue that the coexistence of siderite and magnetite
(\magnetite) in Archean banded iron formations constrain the
atmospheric carbon dioxide concentration to only about 3 times the
present-day level \citep[see also][]{Melnik1982, Kasting2010}. Note,
however, that there is some controversy whether the formation of these
minerals occurred in thermodynamic equilibrium with the
atmosphere--ocean system \citep{Dauphas2011, Reinhard2011,
  Rosing2011}. Indeed it is likely that a considerable fraction of
these minerals were formed during diagenesis rather than in the
supernatant water column and that the conversion of magnetite to
siderite was limited by the rate of supply of organic matter rather
than \cotwo. The simultaneous occurrence of siderite and magnetite in
banded iron formations might thus not provide any strong constraints
on atmospheric \cotwo\ partial pressure after all. The results
published in \citet{Rosing2010} are also in conflict with lower limits
derived from weathering rinds on 3.2-Gyr-old river gravels for which
the presence of iron-rich carbonates requires \cotwo\ partial
pressures of about 10 times pre-industrial levels for the same
environmental temperature of 298~K as in the \citet{Rosing2010} study
\citep{Hessler2004}.

Despite the uncertainties discussed above, geochemical data therefore
suggest that \cotwo\ partial pressures were likely smaller than a few
hundred times pre-industrial levels in the late Archean and early
Proterozoic, meaning that carbon dioxide alone would most likely have
been unable to provide enough warming during these times (see again
Figure~\ref{f:archeanco2}). In this context it should be kept in mind
that all modeling studies which determine the \cotwo\ limit necessary
to warm the early Earth rely on one-dimensional models with highly
parametrized descriptions of many important feedback mechanisms like
the ice-albedo feedback. A further complication arises from
uncertainties in radiative transfer calculations for atmospheres rich
in carbon dioxide \citep{Halevy2009, Wordsworth2010}. The problem
arises because the wings of absorption line profiles and the
parameters governing the continuum absorption of \cotwo\ are poorly
constrained by empirical data for the high \cotwo\ partial pressures
used in calculations of the faint young Sun
problem. \citet{Wordsworth2010}, for example, suggest that the
radiative transfer calculations used in many earlier studies
overestimate the \cotwo\ absorption in the early atmosphere when
compared to a parametrization which most accurately reflects presently
available data.

It therefore remains to be seen whether carbon dioxide concentrations
in agreement with geochemical evidence are sufficient to offset the
faint young Sun.

\subsection{Other greenhouse gases}

Other greenhouse gases have been suggested to contribute to warming
early Earth. For example, ethane (C$_2$H$_6$) is expected to form in
an atmosphere containing methane and exposed to ultraviolet radiation
\citep{Haqq-Misra2008}. It has been shown that ethane can contribute
to an Archean greenhouse \citep{Haqq-Misra2008}, although the effect
is not large as can be seen in Figure~\ref{f:archeanch4}. Warming by
nitrous oxide (N$_2$O) has been suggested \citep{Buick2007}, but
N$_2$O is rapidly photodissociated in the absence of atmospheric
oxygen \citep{Roberson2011}, making it an unviable option for the
Archean.  Furthermore, carbonyl sulfide (OCS) at ppmv levels has the
potential to offset the faint young Sun \citep{Ueno2009}, but it
appears very unlikely that OCS concentrations higher than ppbv (parts
per billion volume) level could have been maintained due to
photodissociation losses \citep{Domagal-Goldman2011}.

Although nitrogen is not a greenhouse gas in itself, a higher partial
pressure of atmospheric nitrogen during the Archean would amplify the
greenhouse impact of other gases by broadening of absorption lines
\citep{Goldblatt2009}. Despite the fact that this additional warming
is partly compensated by increased Rayleigh scattering of short-wave
radiation \citep{Halevy2009}, model calculations show that it could
cause a warming by 4.4$^\circ$C for a doubling of the N$_2$
concentration \citep{Goldblatt2009}. Nitrogen outgassed quickly on
early Earth, so the atmospheric nitrogen content likely equaled at
least the present-day value. Since all nitrogen in the mantle today
must have been processed through the atmosphere, the reservoirs in the
crust and mantle appear sufficiently large to explain higher
atmospheric concentrations and thus a warmer Archean
\citep{Goldblatt2009}.

\subsection{Summary}

In summary, an enhanced greenhouse effect arguably still seems the
most likely solution to the faint young Sun problem. Carbon dioxide
and methane are the most obvious candidates, although they could face
severe difficulties in terms of geochemical constraints and low
production rates, respectively, and their respective
contribution remains uncertain. Ammonia appears less likely than
\cotwo\ and \methane\ because it would have to be shielded against
photodissociation by ultraviolet radiation and because it would be
washed out by rain.

A final assessment of greenhouse-gas warming in the early atmosphere,
however, is complicated by uncertainties in the radiative transfer
functions and the lack of spatially-resolved and fully coupled climate
models for the early Earth comprising the full range of feedbacks in
the Earth system. Finally, other climatic factors like changes in
cloud cover could in principle at least have contributed to a warming
of the Archean Earth.

\section{Clouds in the Archean \\ Atmosphere}
\label{s:clouds}

Clouds exhibit two competing effects on the climate. On the one hand,
clouds, and in particular low clouds, reflect solar radiation back
into space, thus increasing the albedo and cooling the climate. On the
other hand, the water vapor within the clouds absorbs and re-emits
long-wave radiation from the surface and hence warms the planet [see
e.g., \citeauthor{Schneider1972}, 1972, and references therein, as
well as \citeauthor{Stephens2005}, 2005, for a recent review].

The warming effect of a decreased cloud cover (resulting in a lower
albedo and hence an increase in absorbed solar radiation) on the early
atmosphere has been suggested as a possible offset to the faint young
Sun as part of a negative feedback loop in which lower temperatures
decrease (low-level) cloudiness due to a reduction in convective
heating and thus increase the amount of absorbed solar radiation,
counteracting the initial cooling \citep{Henderson-Sellers1979,
  Rossow1982}. This hypothesis has been considered an unlikely
solution for the faint young Sun problem for a long time, however,
because the early Earth was believed to be even warmer than today
(presumably resulting in a higher cloud cover due to increased
evaporation and thus higher reflectivity of the atmosphere), although
more recent studies indicate a more temperate Archean climate (see the
discussion in Section~\ref{s:temphist}). In any case, the precise effect
of cloud feedback for warming or cooling the early Earth remains
uncertain. More recently, \citet{Rosing2010} argued that the Archean
was characterized by larger cloud droplets and shorter cloud
lifetimes, effectively lowering the planetary albedo. Their argument
is based on the presumption that the majority of cloud condensation
nuclei is composed of biologically produced dimethyl sulfide (DMS,
(CH3)$_2$S) and that DMS is produced by eukaryotes only. Both these
assumptions have been challenged, however \citep{Goldblatt2011b}.

It has been hypothesized that a decrease in the cosmic-ray flux due to
the stronger solar wind of the young Sun would decrease cloudiness and
thus provide additional warming to early Earth \citep{Shaviv2003}. For
the present-day climate, the cosmic-ray hypothesis could not be
verified using satellite observations of cloud cover, however
\citep[e.g.,][]{Kristjansson2008, Gray2010}.

The most comprehensive assessment of the effects of clouds on the
early Earth's climate has recently been undertaken by
\citet{Goldblatt2011}.  They find that removing all low clouds (which
increase the albedo, but not the greenhouse effect) yields a forcing
of $\Delta F = 25$~W~m$^{-2}$ and thus only about half the climate
forcing required to offset the faint early Sun ($\Delta F \approx
60$~W~m$^{-2}$ and $\Delta F \approx 40$~W~m$^{-2}$ for the early and
late Archean, respectively), while more realistic reductions of low
cloud cover result in forcings of $\Delta F = 10-15$~W~m$^{-2}$.

In contrast to a diminished cooling effect of low clouds, a stronger
warming due to more thin, high clouds could also contribute to a
warming of the Archean atmosphere. Indeed, such an effect has been
investigated in the context of climate models of an ozone-free
atmosphere \citep{Jenkins1995a, Jenkins1995b, Jenkins1999}. Both
photochemical models \citep{Kasting1979} and the discovery of
mass-independent fractionation (MIF) of sulfur isotopes in rocks older
than 2.45~Ga \citep{Farquhar2000} suggest that the oxygen
concentration in the early atmosphere was very low until $2.3-2.4$~Ga
\citep{Pavlov2002, Bekker2004}, and the Earth hence lacked an ozone
layer. In model experiments, removal of ozone (under present-day
boundary conditions) yields a warming of 2$^\circ$C globally due to an
increase in long-wave cloud radiative forcing \citep{Jenkins1995a,
  Jenkins1995b, Jenkins1999}. The increase in warming in these
simulations is due to the lower temperature in the upper troposphere
and lower stratosphere, leading to higher relative humidity and thus
increased high cloud cover, in particular in higher latitudes.

\citet{Rondanelli2010} focus on the warming effect of high clouds as
well, and suggest that thin cirrus clouds in the tropics could be
sufficient to offset the low solar luminosity. This hypothesis is
based on the `iris mechanism' \citep{Lindzen2001} suggesting a
decrease of tropical cirrus clouds with increasing temperature,
effectively a negative feedback in the present-day Earth system. This
hypothetical mechanism has been extensively challenged in the
literature since no evidence for such an effect could be found in
several satellite data sets \citep{Chambers2002, Fu2002, Hartmann2002,
  Lin2002}.

Independent of the question whether the \citet{Rondanelli2010}
hypothesis appears likely, its effects on the energy balance can be
investigated to estimate its potential importance. Similar to their
assessment of the warming by a decreased low cloud cover,
\citet{Goldblatt2011} find that compensating for the reduced solar
luminosity by enhancing high cloud cover (which adds to the greenhouse
effect) is only possible with full cover of high clouds which are
unrealistically thick and cold. Offsetting the faint young Sun would
require climate forcings of $\Delta F \approx 60$~W~m$^{-2}$ and
$\Delta F \approx 40$~W~m$^{-2}$ for the early and late Archean,
respectively. High clouds can provide a forcing of $\Delta F =
50$~W~m$^{-2}$ if they cover the whole globe and are made 3.5 times
thicker and 14~K colder than conventional wisdom suggests. More
realistic forcings from high clouds during the Archean are estimated
to be 15~W~m$^{-2}$ only and thus insufficient to offset the lower
solar luminosity \citep{Goldblatt2011}.

Hence it appears unlikely that any cloud effect alone can resolve the
faint young Sun problem, although their feedback -- positive or
negative -- certainly plays an important role and should be considered
in any assessment of the Archean climate. The same is true for other
factors influencing the climate on early Earth like its faster
rotation and (potentially) smaller continental area.

\section{Rotational and Continental \\ Effects on Early Earth}
\label{s:rotcont}

\subsection{Rotation and Obliquity}

In modern times, Earth rotates once every $\simeq 24$ hours around its
axis which is tilted at $\simeq 23.5^\circ$ against the ecliptic
(Earth's orbital plane). Although neither variations of the axial tilt
(obliquity) nor of the rotation period directly affect the global
energy balance of the climate system, they can, in principle, change
the distribution of energy within the system. This has effects for the
extent and distribution of ice cover, with consequences for ice-free
regions in the oceans, for Earth's albedo and thus indirectly for the
global energy balance.

High obliquity has been shown to yield a warmer climate and could
offset the faint early Sun for axial tilt values of $65-70^\circ$ in
simulations with an atmospheric general circulation coupled to a slab
ocean for an idealized supercontinent configuration
\citep{Jenkins2000}. At high obliquities, the annual insolation at the
poles is strongly increased ($\sim 220$~W~m$^{-2}$ for an obliquity of
70$^\circ$ and a solar constant reduced by 6\%). Although insolation
at the equator is lowered by $\sim 100$~W~m$^{-2}$ at the same time,
this change in the distribution of insolation is sufficient to prevent
early Earth from global glaciation in these simulations. Paleomagnetic
studies, however, indicate a remarkable stability of Earth's (low)
obliquity over the last 2.5~Gyr \citep{Evans2006}, and it has been
shown that the presence of the Moon stabilizes the obliquity
\citep{Laskar1993}. Even without the Moon, modeled obliquities remain
in a narrow range around the present-day value \citep{Lissauer2012},
suggesting a low obliquity not too different from the present value
since the formation of Earth.

\begin{figure*}
\centerline{\includegraphics[width=10cm]{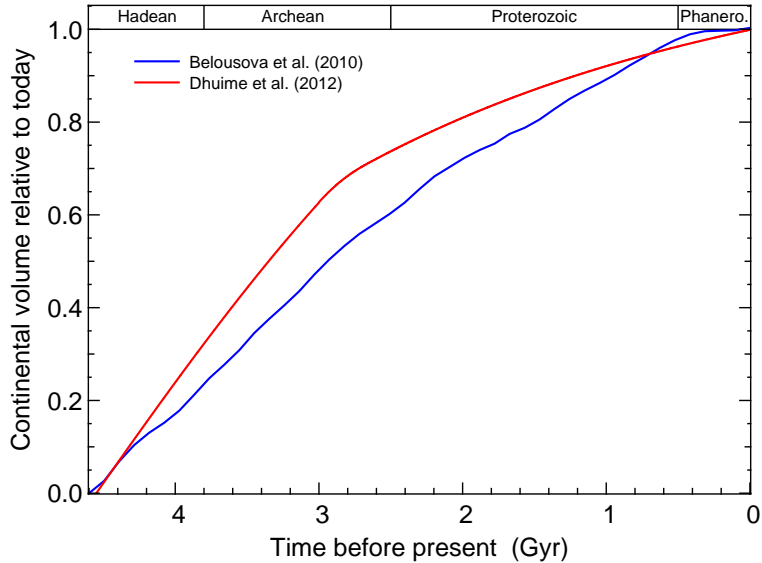}}
\caption{Examples for recent results on the growth of the volume of
  continental crust over time derived from isotopic data
  \citep{Belousova2010, Dhuime2012}.}
\label{f:crust}
\end{figure*}

Tidal friction causes Earth's rotation to slow down and the Moon to
move further away from Earth over time
\citep[e.g.,][]{Williams2000}. For example, Earth's rotation period at
4~Ga has been estimated to be just 14~hours \citep{Zahnle1987}. Using
a simple one-dimensional (zonally averaged) energy balance model to
estimate the effects of a shorter day-length on climate,
\citet{Kuhn1989} find that the effect is important for the Precambrian
climate since it increases the temperature gradient between equator
and poles. This is due to the fact that mid-latitude eddies which are
mostly responsible for the heat transport strongly depend on rotation
rate: at faster rotation rates, these eddies become smaller in size
and thus less efficient in transporting heat polewards. It has been
shown that the rate of meridional heat transport is proportional to
$1/f^2$ \citep{Stone1972}, where $f = 2 \, \Omega \, \sin \phi$ is the
Coriolis parameter depending on Earth's rotation rate $\Omega$ and
latitude $\phi$. This effect could, in principle, prevent low-latitude
glaciation. Note, however, that there is a runaway effect associated
with ice-albedo feedback which pushes the planet into a ``Snowball
Earth'' regime once about half of its surface is covered with ice, see
the discussion below.

Studies using an atmospheric general circulation model coupled to a
simple ocean without heat capacity suggested that fast rotation could
decrease global cloud cover by about 20\% for a day length of 14~h and
thus result in a rise of the global mean air temperature of 2~K
\citep{Jenkins1993a, Jenkins1993b}. In these model experiments, the
decrease in cloudiness is due to a weaker Hadley cell and thus reduced
convection and cloud formation in equatorial latitudes and larger
subsidence in mid-latitudes again reducing cloud cover. A follow-up
study with fixed sea-surface temperatures failed to show the effect,
however, and found a small increase in global cloud cover
\citep{Jenkins1996}.

Sensitivity studies carried out with atmospheric general circulation
models for different rotation periods demonstrate the importance of
the rotation rate for the structure and strength of the atmospheric
circulation \citep{Williams1988, Navarra2002}: with increasing
rotation rate, the Hadley and Ferrel cells become generally narrower
and weaker, the polar cell tends to split into smaller cells, and the
temperature gradient between the poles and the equator increases. How
these changes interact with the ocean, however, has still to be
demonstrated with fully coupled models using a general-circulation
ocean module.

\subsection{Continental area}

A further striking difference between the Archean world and the
present-day Earth is the fraction of the surface covered by
continents. During the Archean, the land area has been estimated to
comprise only about 10\% of today's continental area
\citep{Goodwin1981}. Earlier models for continental growth yielded
widely diverging growth curves for continental volume \citep[see,
e.g.,][for an overview]{Kroener1985, Flament2009}, but recent work
\citep{Belousova2010, Dhuime2012} based on the isotopic composition of
zircons provides much better constraints on the evolution of
continental volume, which is illustrated in
Figure~\ref{f:crust}. While continental volume has grown to $\sim
70$\% by the end of the Archean, it appears likely that a smaller
fraction of Earth's surface was covered by land during the early
Archean, which affected both the albedo and heat transport processes
in the Earth system.

The lower albedo due to the smaller continental area has been
suggested several times as an important factor for the energy budget
of the Archean climate \citep{Schatten1982, Cogley1984, Gerard1992,
  Jenkins1993a, Molnar1995, Rosing2010}. It can be easily shown,
however, that the effect of a lower surface albedo alone is
insufficient to offset the decrease in solar radiation during the
Archean \citep{Walker1982, Kuhn1989}. \citet{Goldblatt2011b} estimate
that the decreased surface albedo cannot contribute more than
5~W~m$^{-2}$ in radiative forcing to any solution of the faint young
Sun problem, much less than the values of $\Delta F \approx
60$~W~m$^{-2}$ and $\Delta F \approx 40$~W~m$^{-2}$ required during
the early and late Archean, respectively. Despite a decrease in
surface albedo, some studies have even suggested an \textit{increase}
in global albedo under global-ocean conditions due to higher cloud
fractions caused by increased evaporation, although the results
strongly depend on the amount of heat transported from low to high
latitudes which has been prescribed in these simulations
\citep{Jenkins1995a, Jenkins1995b, Jenkins1999}.

\begin{figure*}[t]
\centerline{\includegraphics[width=10cm]{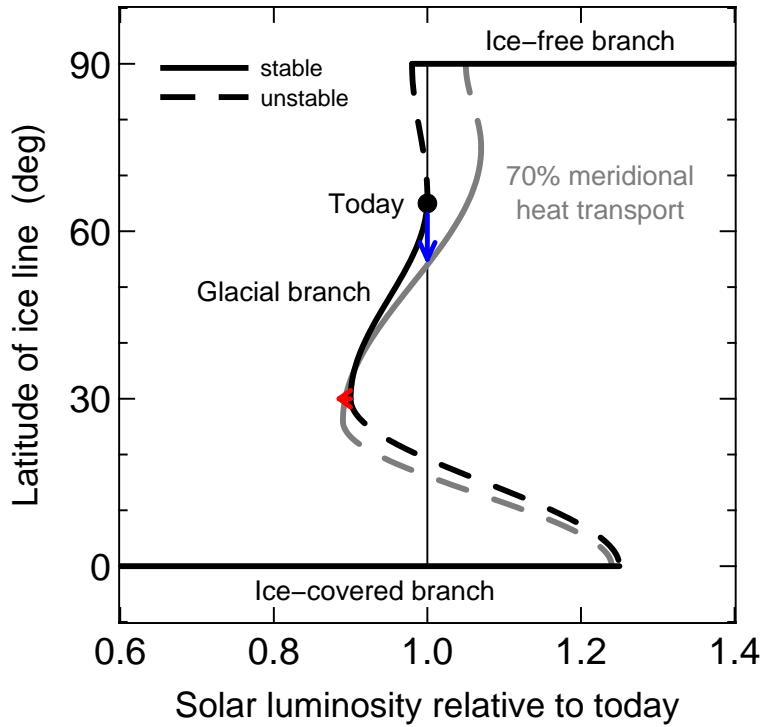}}
\caption{Schematic diagram illustrating the position of the ice line
  as a function of solar luminosity. The positive ice-albedo feedback
  results in an instability leading to run-away glaciation once the
  ice line reaches $\sim 30^\circ$ in latitude. Stable branches are
  indicated by \textit{solid}, unstable branches by \textit{dashed
    lines}. The situation for a reduced meridional heat transport is
  shown in \textit{gray}. The \textit{blue arrow} illustrates the
  change in the location of the iceline at a given solar luminosity,
  while the \textit{red arrow} shows the associated change in minimum
  solar luminosity for the stable glacial branch. Modified after
  \citet{Ikeda1999} and \citet{Hoffman2002}.}
\label{f:iceline}
\end{figure*}

In addition to the lower surface albedo, the smaller continental area
could have a substantial effect on the heat transport in the Archean
oceans and thus the extent of polar ice caps. The influence of
meridional heat transport on the latitude of the ice line is
illustrated in Figure~\ref{f:iceline}, which is based on results from
simple energy balance models and assumptions about albedo changes
\citep{Ikeda1999}. A reduced meridional heat transport indeed results
in the ice line being located closer to the equator for a given solar
luminosity (or greenhouse-gas concentration) in the stable regime with
existing polar caps as indicated by the blue arrow in the Figure. The
lower limit in solar luminosity beyond which this stable branch can be
occupied, on the other hand, is only slightly affected by meridional
transport, see the red arrow in the Figure. Again, these effects would
have to be verified with more comprehensive and spatially resolved
models to explore the sensitivities of the ice line on geography
\citep{Crowley1993}, the dynamics of sea ice \citep{Hyde2000} and
ocean dynamics \citep{Poulsen2001}.

\citet{Endal1982} suggested that the smaller land fraction in the
Archean might have intensified the meridional heat transport in the
oceans, thus pushing the boundary of polar ice caps towards higher
latitudes. Naively one would expect, however, that the absence of land
barriers would lead to a predominantly zonal ocean circulation with
reduced heat transport to the polar regions.  Indeed, later studies
with improved (but still comparatively simple) ocean models found a
weak meridional heat transport and thus large temperature gradients
between the equator and the poles \citep{Henderson-Sellers1988,
  Longdoz1997}. The same behavior was found in simulations with
state-of-the-art general circulation models for a planet without any
landmass, an ``aquaplanet'' \citep{Marshall2007, Enderton2009,
  Ferreira2010}.

Note that the extent of exposed continental area and its geographic
distribution, where the latter is essentially unknown for the Archean,
also affect chemical weathering and thus the carbon cycle
\citep{Marshall1988}.

In addition to the rotation rate and the continental distribution,
there are other important differences between the Archean and the
modern ocean like its possibly higher salinity and increased mixing
due to tidal activity, which should be taken into account in future
studies of Archean ocean circulation.

\subsection{Ocean salinity and tides}

Ocean salinity is certainly an important environmental variable, but
its evolution over Earth's history is difficult to reconstruct
\citep{Hay2006}. Based on massive salt beds deposited in more recent
times, it has been suggested that the Archean ocean was characterized
by a salinity at least a factor of 1.6 higher than today
\citep{Knauth2005}. In today's ocean, salinity is dominated by sodium
chlorite (NaCl), and it is likely that this was also the case in the
Archean. Chlorine (Cl$^-$) outgassed early in the history of life and
was dissolved in the earliest ocean \citep{Holland1984}. Initial
salinity can be estimated from the volume of salt now deposited in
massive salt beds and subsurface brines, resulting in salinities up to
a factor of 2 higher than today. This appears convincing at least for
the early Archean; whether it also holds for the late Archean depends
on assumptions about continental evolution during the Archean
eon. Unfortunately, geochemical constraints on Archean ocean salinity
are currently missing. \citet{deRonde1997} analyzed fluid inclusions
in 3.2~Ga deposits interpreted as mid-Archean hydrothermal vents and
found chlorine concentrations 1.65 larger than today, but these
formations have later been re-interpreted as Quaternary spring
deposits \citep{Lowe2003}.

Total ocean salinity is of interest for the Archean climate system
because it could in principle influence the thermohaline ocean
circulation \citep{Kuhlbrodt2007}. Indeed, \citet{Hay2006} observed
that for present-day salinities of $\simeq 35$\permil\ the density of
seawater changes only weakly with temperature when approaching
freezing point, requiring an enhancement of the salt content by
sea-ice formation or evaporation to make seawater dense enough to sink
to the ocean's interior.  For an ocean with salinities above $\simeq
40$\permil, the density increases with falling temperature. Therefore,
an energy-consuming phase transition during deep-water formation would
not be required, which could yield a stronger thermohaline
circulation. This claim has been refuted in a modeling experiment by
\citet{Williams2010}, however, which in fact shows a weaker meridional
overturning in an ocean model with twice the present-day salinity and
today's topography. The physical reasons for these conflicting
assessments of circulation strength for higher average salinity remain
unclear, however, and the influence of global salinity on ocean
circulation certainly merits further study.

Tidal activity was higher during the Archean due to the smaller orbit
of the Moon which affects mixing in the ocean and thus, in turn, ocean
circulation and marine heat transport \citep{Munk1998}. Both effects
should be explored in more detail with state-of-the-art ocean general
circulation models.

\subsection{Summary}

Rotational and continental effects are thus important for assessing
the warming effects on the Archean climate. It is likely that they
cannot solve the faint young Sun problem on their own, for which an
enhanced greenhouse effect on early Earth appears to be required. The
influences of faster rotation and different continental configuration,
however, are important for understanding the energy budget and
dynamics of the Archean climate system, so any convincing
demonstrations of solutions involving enhanced levels of greenhouse
gases will require simulations with fully coupled state-of-the-art
climate models including these effects.

\section{Conclusions and Future \\ Directions}
\label{s:disc}

After four decades of research the faint young Sun problem indeed
``refuses to go away'' \citep{Kasting2010}. To a large extent, this is
certainly due to the still limited knowledge of the conditions on
early Earth, although the last decades have seen considerable
progress, and some parameters are now better constrained than they
used to be in the past. Nevertheless, improved constraints on
atmospheric composition during the Archean eon would obviously be
extremely important, although certainly challenging to obtain. Despite
the difficulties involved, there have certainly been remarkable
advances in geochemistry in recent years. Note, for example, that most
of the geochemical constraints on Archean and Proterozoic carbon
dioxide partial pressures shown in Figure~\ref{f:archeanco2} were
derived within the last decade. There is thus reason to be hopeful in
continued progress in this area.

In addition to better data, however, improvements in the efforts on
modeling the Earth's climate during the Archean are urgently needed --
as on other important problems in deep-time paleoclimatology like
climate changes associated with mass-extinction events
\citep{Feulner2009} or greenhouse climates of the past
\citep{Huber1999}. Many suggested solutions to the faint young
Sun problem, especially those involving continental or albedo effects,
require spatially resolved climate simulations rather than the
one-dimensional or simple energy balance atmospheric models
traditionally used in studies of the faint young Sun problem, and full
coupling to state-of-the-art ocean and sea-ice models. Finally, the
full range of feedback mechanisms has to be explored in detail.

There are several challenges in all modeling efforts of the Archean
climate. First, there are still considerable uncertainties in key
climate characteristics like greenhouse-gas concentrations or
continental configuration. These parameter uncertainties have to be
properly quantified using ensemble simulations of the Archean climate
system.  Because of their higher speed, this is traditionally the
domain of intermediate-complexity climate models \citep{Claussen2002}.

Secondly, essentially all of the more comprehensive climate models are
to some extent tuned to present-day climate conditions. To be able to
apply them to the early Earth's climate and obtain meaningful results,
they have to provide robust results for a climate state which is
considerably different than today. Not only for this reason, the
emphasis in all climate modeling efforts for the faint young Sun
problem should lie in improving our understanding of the physical
processes characterizing the Archean climate system. Finally it would
be advisable to simulate the Archean climate with several models using
different approaches to be able to compare model results.

Given the continued interest this important topic enjoys, the next
decade might bring us closer to finally answering the question of how
water on early Earth could have remained liquid under a faint young
Sun, certainly one of the most fundamental questions in
paleoclimatology.



%
%
%
%
%
%

%
%
\section*{Glossary}

\paragraph{Albedo} Reflectivity of a planet, defined as the ratio of
reflected to incoming radiation.

\paragraph{Anti-greenhouse effect} Effect of atmospheric gases which
are opaque for incoming solar radiation but allow thermal radiation
from the surface to escape to space.

\paragraph{Aquaplanet} Idealized planet fully covered by an ocean.

\paragraph{Archean} Geological eon lasting from $3.8 \times 10^9$ to
$2.5 \times 10^9$ years ago.

\paragraph{Banded iron formation} Sedimentary rock consisting of
alternating layers of iron oxides and iron-poor rock.

\paragraph{Bolometric luminosity}
Luminosity (radiative energy emitted per unit time) integrated over
all wavelengths.

\paragraph{Cosmic rays} High-energy charged particles (mostly protons,
helium and heavier nuclei, electrons) reaching Earth's atmosphere from
space.

\paragraph{Diagenesis} Sum of all (mostly chemical) low-temperature
and low-pressured processes by which sediments are altered after
deposition but before conversion to rock (lithification).

\paragraph{Ecliptic} Earth's orbital plane.

\paragraph{Exosphere} The uppermost layer of Earth's atmosphere.

\paragraph{Ferrel cells} Meridional atmospheric circulation pattern
between the Hadley and the polar cells.

\paragraph{Hadean} Geological eon lasting from the formation of the
Earth $4.56 \times 10^9$ years ago to the beginning of the Archean
$3.8 \times 10^9$ years ago.

\paragraph{Hadley cells} Tropical part of the meridional atmospheric
circulation, with rising air near the equator, poleward motion in the
upper troposphere, sinking air in the subtropics (around 30$^\circ$
latitude in the present-day climate) and a surface flow towards the
equator.

\paragraph{Helioseismology} Technique to gain insight into the Sun's
interior structure from observations of resonant oscillations at the
solar surface.

\paragraph{Hydrothermal vent} Source of water heated by contact with
hot magma in volcanically active areas, commonly used to describe hot
springs on the ocean floor.

\paragraph{Late Heavy Bombardment} Period of intense collision of
asteroids and comets with solar-system planets and moons inferred from
a spike in lunar cratering rates $\sim 3.9 \times 10^9$ years ago.

\paragraph{Magnetosphere} Region of interaction between Earth's
intrinsic magnetic field and the stream of charged particles from the
Sun (the solar wind).

\paragraph{Main sequence} Historically identified as a well-defined
band in a color-brightness diagram of stars, the main sequence period
is the time in the life of a star during which it generates energy by
nuclear fusion of hydrogen to helium in its core.

\paragraph{Mesoproterozoic} Geological era in the Proterozoic lasting
from $1.6 \times 10^9$ years ago to $1.0 \times 10^9$ years ago.

\paragraph{Methanogenic bacteria (methanogens)} Group of anaerobic
microorganisms which produce methane.

\paragraph{Obliquity} Tilt of Earth's rotation axis against its
orbital plane.

\paragraph{Paleoproterozoic} Earliest geological era within the
Proterozoic eon spanning the time from $2.5$ to $1.6 \times 10^9$
years ago.

\paragraph{Paleosol} Layer of fossilized soil.

\paragraph{Photolysis, photodissociation} Destruction of a chemical
compound by photons.

\paragraph{Planetesimals} Solid objects with sizes of one kilometer
and larger forming in the rotating disk around young stars.

\paragraph{Polar cells} High-latitude atmospheric circulation pattern
similar to the Hadley cells, with rising air around 60$^\circ$
latitude in the present-day climate, poleward motion in the upper
troposphere, descending air around the poles, and a surface flow
towards the equator to close the loop.

\paragraph{Precambrian} Informal name for the geological time before
the Cambrian, i.e., older than $542 \times 10^6$ years ago.

\paragraph{Primordial nucleosynthesis} Formation of atomic nuclei
beyond light hydrogen ($^1$H) shortly after the big bang, resulting in
the production of the stable nuclei of deuterium ($^2$H), the helium
isotopes $^3$He and $^4$He and the lithium isotopes $^6$Li and $^7$Li.

\paragraph{Proterozoic} Geological eon lasting from the end of the
Archean $2.5 \times 10^9$ years ago to $542 \times 10^6$ years ago.

\paragraph{Protoplanetary disk} Rotating disk of dense gas and dust
surrounding a newly formed star.

\paragraph{Quaternary} Geological period spanning the last $2.6 \times
10^6$ years.

\paragraph{Radiative forcing} Change in net irradiance (downwards
minus upwards) at the upper limit of the troposphere, thus
characterizing changes in the energy budget of the surface-troposphere
system.

\paragraph{Salinity} Measure of the dissolved salt content of ocean
water, usually expressed as parts per thousand.

\paragraph{Solar analogs} Stars with physical and chemical
characteristics similar to the Sun.

\paragraph{Solar constant} Total radiative energy per unit time and
unit area incident on a plane perpendicular to the direction to the
Sun and at the mean distance between Sun and Earth.

\paragraph{Solar luminosity} Radiative energy per unit time emitted by
the Sun.

\paragraph{Solar wind} Stream of charged particles (mostly electrons
and protons) originating in the Sun's upper atmosphere.

\paragraph{Standard solar model} Numerical model of the structure and
evolution of the Sun based on fundamental equations of stellar physics
and constrained by the observed physical and chemical characteristics
of the present-day Sun.

\paragraph{Stromatolites} Lithified, sedimentary structures growing
via sediment trapping by microbial mats.

\paragraph{Supernatant} The supernatant water column is the water
overlying sedimented material.

\paragraph{Thermohaline circulation} Large-scale ocean currents driven
by density gradients due to heat and freshwater fluxes at the ocean
surface.

\paragraph{Troposphere} The lowermost layer of Earth's atmosphere.

\paragraph{Zero-age main sequence} Position of stars in a
brightness-color diagram which have just started nuclear fusion of
hydrogen to helium in their cores.

%
\begin{notation}
$a$  & semi-major axis of Earth's elliptical orbit \\
$A$ & albedo \\
\methane\ & methane \\
C$_2$H$_6$ & ethane \\
CaCO$_3$ & calcium carbonate \\
CaSiO$_3$ & wollastonite \\
CaSO$_4$ & anhydrite \\
CaSO$_4\cdot$H$_2$O & gypsum \\
CH$_2$O  & formaldehyde \\
(CH3)$_2$S & dimethyl sulfide \\
Cl$^-$ & chlorine \\
\cotwo\ & carbon dioxide \\
$\delta^{18}O$ & measure of the ratio of the stable oxygen isotopes
$^{18}$O and $^{16}$O \\
$f$ & Coriolis parameter \\
$\Delta F$ & radiative forcing \\
$\varepsilon$ & surface emissivity \\
\siderite\ & siderite \\
\magnetite\ & magnetite \\
$G$ & gravitational constant \\
HCN & hydrogen cyanide \\
\water\ & water \\
\htwos\ & hydrogen sulfide \\
$L$ & bolometric solar luminosity as a function of time \\
$L_\odot$ & present-day bolometric solar luminosity \\
$M$ & solar mass as a function of time \\
$\dot{M}$ & solar mass-loss rate \\
$\dot{M}_\mathrm{fusion}$ & rate of solar mass loss due to nuclear
fusion \\
$\dot{M}_\mathrm{wind}$ & rate of solar mass loss due to solar wind \\
$M_\odot$ & present-day solar mass \\
\ntwo\ & molecular nitrogen \\
\amm\ & ammonia \\
NaCl & sodium chlorite \\
N$_2$O & nitrous oxide \\
\otwo\ & molecular oxygen \\
OCS & carbonyl sulfide \\
OH & hydroxyl \\
$\Omega$ & Earth's rotation rate\\
$\Omega_\odot$ & solar rotation rate \\
$\phi$ & geographic latitude \\
$r$ & mean distance between Sun and Earth \\
$R$ & radius of the Earth \\
$S_0$ & solar constant \\
$\sigma$ & Stefan-Boltzmann constant \\
$t$ & time \\
$t_\odot$ & age of the Sun \\
$T_s$ & surface temperature \\
$\tau^\ast$ & column infrared gray opacity
\end{notation}
%

\begin{acknowledgments}
  It is a pleasure to thank the two reviewers, Colin Goldblatt and
  James Kasting, as well as the editor, Mark Moldwin, for their
  comments which helped to improve this review paper
  considerably. Furthermore, I would like to thank Hendrik Kienert for
  numerous discussions and helpful comments on earlier drafts of the
  manuscript. I am grateful to Alison Schlums for proofreading and to
  Grit Steinh\"ofel-Sasgen for background information on banded iron
  formations. This research has made use of NASA's Astrophysics Data
  System Bibliographic Services.
\end{acknowledgments}

\end{article}




%
%
%
%
%
%


\end{document}